\documentclass[12pt]{article}

\usepackage{epsfig}
\usepackage{amstex}

\newcommand{\ho}{\bar{h}}
\newcommand{\hu}{\mbox{\b{$h$}}}

\newcommand{\dxcov}{dx_{\mbox{\scriptsize cov}}}
\newcommand{\dtri}{d^{\,3}}
  
\newcommand{\cla}{{\mathcal A}}
\newcommand{\clb}{{\mathcal B}}
\newcommand{\cld}{{\mathcal D}}
\newcommand{\cle}{{\mathcal E}}
\newcommand{\cli}{{\mathcal I}}

\newcommand{\cll}{{\mathcal L}}

\newcommand{\cls}{{\mathcal S}}
\newcommand{\clt}{{\mathcal T}}
\newcommand{\clw}{{\mathcal W}}
\newcommand{\clx}{{\mathcal X}}
\newcommand{\clz}{{\mathcal Z}}
\newcommand{\bk}{\mbox{\boldmath $k$}}

\newcommand{\kev}{\mbox{\rm \,keV}}

\newcommand{\kmsmpc}{\mbox{$\mathrm{kms^{-1}\,Mpc^{-1}}$}}

\newcommand{\ord}{{\mathrm O}}
\newcommand{\simgt}{\,\raisebox{-0.6ex}{$\stackrel{\textstyle{>}}{\sim}$}\,}
\newcommand{\simlt}{\,\raisebox{-0.6ex}{$\stackrel{\textstyle{<}}{\sim}$}\,}
 
\newcommand{\cldh}{\widehat{{\cal D}}}
\newcommand{\cldb}{\bar{{\cal D}}}
\newcommand{\cldhsq}{{\cldh}^{2}}

\newcommand{\qk}{Q_{k}}
\newcommand{\qkv}{Q_{k}^{(v)}}

\newcommand{\qkl}[1]{\qk^{(#1)}}
\newcommand{\qkldt}[1]{\dot{Q}_{k}^{(#1)}}
 
\newcommand{\fgam}{f_{\gamma}}
\newcommand{\fnu}{f_{\nu}}
\newcommand{\jgaml}[1]{J^{(#1)}_{\gamma}}
\newcommand{\jnul}[1]{J^{(#1)}_{\nu}}
\newcommand{\fgaml}[1]{F^{(#1)}_{\gamma}}

\newcommand{\fgamldt}[1]{\dot{F}^{(#1)}_{\gamma}}

\newcommand{\jgamldt}[1]{\dot{J}^{(#1)}_{\gamma}}
\newcommand{\jnuldt}[1]{\dot{J}^{(#1)}_{\nu}}
\newcommand{\jnuldtk}[1]{\dot{J}^{(#1)}_{\nu k}}
\newcommand{\jgamldtk}[1]{\dot{J}^{(#1)}_{\gamma k}}
\newcommand{\jgamlk}[1]{J^{(#1)}_{\gamma k}}
\newcommand{\jnulk}[1]{J^{(#1)}_{\nu k}}
\newcommand{\qgamk}{q_{\gamma k}}
\newcommand{\qgamdtk}{\dot{q}_{\gamma k}}
\newcommand{\qnuk}{q_{\nu k}}
\newcommand{\qnudtk}{\dot{q}_{\nu k}}
\newcommand{\pigamk}{\pi_{\gamma k}}
\newcommand{\pigamdtk}{\dot{\pi}_{\gamma k}}
\newcommand{\pinuk}{\pi_{\nu k}}
\newcommand{\pinudtk}{\dot{\pi}_{\nu k}}
 \newcommand{\rhob}{\rho_{b}}
\newcommand{\prb}{p_{b}}
\newcommand{\rhogam}{\rho_{\gamma}}
\newcommand{\rhonu}{\rho_{\nu}}
\newcommand{\velb}{v_{b}}
\newcommand{\vk}{v_{k}}
\newcommand{\rhocdm}{\rho_{c}}
\newcommand{\velcdm}{v_{c}}
\newcommand{\clxcdm}{\clx_{c}}
\newcommand{\clxb}{\clx_{b}}
\newcommand{\clxgam}{\clx_{\gamma}}
\newcommand{\clxnu}{\clx_{\nu}}
\newcommand{\clxgamk}{\clx_{\gamma k}}
\newcommand{\clxnuk}{\clx_{\nu k}}
\newcommand{\clxbk}{\clx_{bk}}
\newcommand{\clxcdmk}{\clx_{ck}}
\newcommand{\clxgamdtk}{\dot{\clx}_{\gamma k}}
\newcommand{\clxnudtk}{\dot{\clx}_{\nu k}}
\newcommand{\clxbdtk}{\dot{\clx}_{bk}}
\newcommand{\clxcdmdtk}{\dot{\clx}_{ck}}
 
\newcommand{\dt}{\! \cdot \!}
\newcommand{\wdg}{\! \wedge \!}
\newcommand{\half}{{\textstyle \frac{1}{2}}}

\newcommand{\third}{{\textstyle \frac{1}{3}}}
\newcommand{\qrt}{{\textstyle \frac{1}{4}}}
\newcommand{\da}{\partial_{a}}
\newcommand{\db}{\partial_{b}}

\newcommand{\csound}{c_{s}}

\newcommand{\veqspc}{\vspace{-\belowdisplayskip}\vspace{-0.2cm}\vspace{-\abovedisplayskip}}
\newcommand{\thte}{\theta_{\!e}}

\newcommand{\etal}{\emph{et al.\/}}
\newcommand{\et}[1]{e^{\mbox{\footnotesize $#1$}}}

\numberwithin{equation}{section}

\begin{document}

\begin{center}
{\bf\large RECENT DEVELOPMENTS IN THE CALCULATION OF\\%
CMB ANISOTROPIES}
 
\vspace{0.4cm}
{\large Anthony Challinor and Anthony Lasenby}\\
{\it MRAO, Cavendish Laboratory, Madingley Road,} \\
{\it Cambridge CB3 0HE, UK.}\\
\vspace{0.4cm}
\today
\end{center}
\vspace{0.4cm}

\begin{abstract}
We discuss three recent developments in the calculation of Cosmic Microwave
Background (CMB) anisotropies. We begin with a discussion of the
relativistic corrections to the Sunyaev-Zel'dovich effect. By extending
the Kompaneets equation to include relativistic effects, we are able to
to derive simple analytic forms for the spectral changes
due to the thermal Sunyaev-Zel'dovich effect. Relativistic corrections
result in a small reduction in the amplitude of the effect over the
Rayleigh-Jeans region, which for a typical cluster temperature of $8\kev$,
amounts to a correction downwards to the value of the Hubble constant
derived from combined X-ray and Sunyaev-Zel'dovich information by about 5
percent.
Our second topic is a discussion of covariant kinetic theory methods for
the gauge-invariant calculation of primordial CMB anisotropies.
We present a covariant version of the Boltzmann equation,
which we use to derive a set of covariant equations for
the gauge-invariant variables in a Cold Dark Matter (CDM) model, which are
independent of the background curvature and type of perturbation.
Equations describing a particular type of perturbation
are obtained readily from this set, as we demonstrate for scalar perturbations
in a $K=0$ universe. By integrating the covariant Boltzmann equations along
the line of sight, and using the instantaneous recombination approximation,
we obtain an expression for the large scale anisotropy which
agrees with an expression derived recently by Dunsby.
Finally, we discuss some recent accurate calculations of the CMB anisotropy
in global defect models. In such models, the effect of vorticity generation
by the causal sources proves to be significant, leading to a suppression of
acoustic peaks. The result is that global defect models of
structure formation may already be at variance with the
growing volume of CMB (and large scale structure) data.
\end{abstract}

\section{Introduction}

Future satellite experiments hold the promise of obtaining the first high
resolution, full-sky maps of the CMB. These maps will contain a wealth
of cosmological information, allowing unprecedented accuracy in the
determination of cosmological parameters, and the opportunity to
distinguish between rival theories of structure formation. Interpretation of
the CMB data requires not only accurate calculation of the primordial
anisotropies in specific cosmological models, but also accurate modelling of
sources of secondary anisotropies and foregrounds.
In this lecture we consider three recent developments in the calculation
of primary and secondary anisotropies. We begin, in Section~\ref{sec_sz},
with secondary anisotropies,
considering the effect of relativistic corrections on the spectral distortion
of the CMB from Compton scattering in hot clusters~\cite{chall97-sz}
(the Sunyaev-Zel'dovich effect).
We then turn to primary anisotropies,
presenting a summary of some work in progress~\cite{chall97-cmb}
on gauge-invariant calculations
of the propagation of primordial anisotropies, using covariant kinetic theory
methods, in Section~\ref{sec_kin}.
We use a version of the covariant perturbation
theory developed by Ellis and coworkers (see, for
example, Ellis~\etal~\cite{ellis89a,ellis89b}).
Their approach is superior to others,
including the Bardeen formalism~\cite{bardeen80}, since it deals only with
physically-defined, gauge-invariant variables. We end with a brief review
of accurate calculations by Pen, Seljak and Turok~\cite{pen97,seljak97b}
of the CMB power spectra in global defect theories, in Section~\ref{sec_defs}.
These calculations show that there are problems reconciling global defect
models of structure formation with observation.

We employ natural units $c=\hbar=G=1$, except when presenting numerical
results in Section~\ref{sec_sz}.

\section{Relativistic Corrections to the Sunyaev-\\%
Zel'dovich Effect}
\label{sec_sz}

The Sunyaev-Zel'dovich effect is concerned with the generation of
CMB spectral distortions by inverse Compton scattering in hot clusters.
Non-relativistic treatments of this effect usually employ
the Kompaneets equation~\cite{komp57} to determine the distortion.
However, the Kompaneets equation does not include relativistic effects, which
may be important for hot clusters where $k_{B}T_{e} \simgt 10 \kev$.
For this reason, and because of the low optical depth of typical clusters,
relativistic treatments of the Sunyaev-Zel'dovich effect usually employ
a multiple scattering description of the
\nocite{wright79,fabbri81,taylor89,loeb91,reph95}
Comptonization process~\cite{wright79}--\cite{reph95}.
Including relativistic effects in this procedure gives a complicated expression
for the spectral distortion, which is best handled by numerical
techniques (see, for example, Rephaeli~\cite{reph95}).
 
In this section, we present an extension of the Kompaneets equation which
includes relativistic effects in a self-consistent manner. The relativistic
extension to the Kompaneets equation has been
considered independently by Stebbins~\cite{stebbins96,stebbins97}.
The extended Kompaneets equation allows
the Sunyaev-Zel'dovich effect in hot clusters to be described
analytically on the basis of a Kompaneets type equation, instead of
the previous numerical approaches.
Simple analytic forms are given for the spectral distortions in the
limit of small optical depth, including relativistic effects to
second-order. These are
in excellent agreement with Rephaeli's numerical calculations~\cite{reph95},
which were based on the multiple scattering approach (truncated at one
scattering). This lends further support to Fabbri's observation
that the Boltzmann equation can be applied to describe the Sunyaev-Zel'dovich
effect in optically thin clusters~\cite{fabbri81}, despite claims to the
contrary~\cite{wright79}.
 
The relativistic corrections to the Sunyaev-Zel'dovich effect are
important in the calculation of the Hubble constant $H_{0}$ by
the Sunyaev-Zel'dovich route in hot clusters (see, for
example, Lasenby and Jones~\cite{las97} for a recent review of the
Sunyaev-Zel'dovich route of determining $H_{0}$, and Saunders~\cite{saun96}
for recent observations).
In the Rayleigh-Jeans region, we find that
relativistic effects lead to a small decrease in the Sunyaev-Zel'dovich effect,
and hence a small reduction in the hitherto determined values of $H_{0}$, in
agreement with the conclusions of Rephaeli and Yankovitch~\cite{reph97}.
 
\subsection{Extending the Kompaneets Equation}
\label{sz_sec_komp}

In this lecture we shall not consider effects due to the peculiar motion of the
cluster (such effects give rise to a kinetic correction to the
Sunyaev-Zel'dovich effect). For a comoving cluster, the CMB photon distribution
function is isotropic and may be denoted $n(\omega)$, where $\omega$ is the
photon frequency. The electrons are assumed to be in thermal equilibrium at
temperature $T_{e}$, and are described by an isotropic distribution
function $f(E)$, where $E$ is the electron energy.
The Boltzmann equation
describing the evolution of $n(\omega)$ may be written as~\cite{buch76}
\begin{equation}
\frac{\partial n(\omega)}{\partial t} = -2 \int \frac{\dtri p}{(2\pi)^{3}}
\dtri p' \dtri k' \,W \Bigl[n(\omega)\bigl(1+ n(\omega')\bigr)f(E) -
n(\omega')\bigl(1+n(\omega)\bigr)f(E') \Bigr],
\adjusttag{7pt}
\label{sz_eq_1}
\end{equation}
where $W$ is the invariant transition amplitude for Compton
scattering of a photon of 4-momentum $k^{\mu}$
by an electron (of charge $e$ and mass $m$) with
4-momentum $p^{\mu}$, to a photon momentum $k^{\prime\mu}$ and an
electron momentum $p^{\prime\mu}$~\cite{ber-quan}:
\begin{equation}
\begin{aligned}
W &= \frac{(e^{2}/4\pi)^{2} \bar{X}}{2\omega\omega' E E'}
\delta^{4}(p+k-p'-k') \\
\bar{X} &\equiv 4m^{4} \left(\frac{1}{\kappa} + \frac{1}{\kappa'}\right)^{2} -
4m^{2} \left(\frac{1}{\kappa} + \frac{1}{\kappa'}\right) -
\left(\frac{\kappa}{\kappa'} + \frac{\kappa'}{\kappa}\right),
\end{aligned}
\label{sz_eq_2}
\end{equation}
with $\kappa \equiv -2p^{\mu}k_{\mu}$ and $\kappa'\equiv
2p^{\mu}k^{\prime}_{\mu}$. In equation~\eqref{sz_eq_1},
we have assumed that electron
degeneracy effects may be ignored.

The electrons are described by a relativistic Fermi distribution. Since we
are ignoring degeneracy effects, we have
\begin{equation}
f(E)\approx \et{-(E-\mu)/k_{B}T_{e}}.
\end{equation}
Substituting this form for $f(E)$ into equation~\eqref{sz_eq_1},
and expanding the
term in brackets in the integrand in powers of $\Delta x$, where
\begin{align}
x &\equiv \frac{\omega}{k_{B}T_{e}} \\
\Delta x &\equiv \frac{\omega'-\omega}{k_{B}T_{e}},
\end{align}
gives a Fokker-Planck expansion
\begin{multline}
\frac{\partial n(x)}{\partial t} = 2 \left(\frac{\partial n}{\partial x}
+ n(1+n)\right) I_{1} 
+ 2 \left(\frac{\partial^{2}n}{\partial x^{2}} +
2(1+n)\frac{\partial n}{\partial x} + n(1+n) \right) I_{2} \\
+ 2\left(\frac{\partial^{3}n}{\partial x^{3}} + 3(1+n)
\frac{\partial^{2}n}{\partial x^{2}} + 3 (1+n) 
\frac{\partial n}{\partial x} + n(1+n) \right) I_{3}
\\
+ 2\left(\frac{\partial^{4} n}{\partial x^{4}} + 4(1+n)
\frac{\partial^{3} n}{\partial x^{3}} + 6(1+n)
\frac{\partial^{2} n}{\partial x^{2}} + 4(1+n)
\frac{\partial n}{\partial x} + n(1+n) \right) I_{4} + \cdots,  
\label{sz_eq_3}
\end{multline}
where
\begin{equation}
I_{n} \equiv
\frac{1}{n!}\int \frac{\dtri p}{(2\pi)^{3}} \dtri p' \dtri k' \,W f(E)
(\Delta x)^{n},
\label{sz_eq_4}
\end{equation}
which does not depend on $n(\omega)$.

The calculation of the $I_{n}$ may be performed by
expanding the integrand in powers of $p/m$ and $\omega/m$. Performing the
integral over $E$ requires an asymptotic expansion of the electron
distribution function $f(E)$.
Evaluating the $I_{n}$, we develop a (possibly asymptotic)
expansion of $\partial n/\partial t$ in $\thte$, where
\begin{equation}
\thte \equiv \frac{k_{B}T_{e}}{m}.
\end{equation}
In this lecture, we shall only be
concerned with the lowest order relativistic corrections, and start by
retaining all terms up to $\ord(\thte^{2})$. To include all such terms
consistently, it is necessary to retain only the first four terms in the
series~\eqref{sz_eq_3}. A lengthy calculation gives the result
\begin{equation}
\frac{\partial n(x)}{\partial t} = \sigma_{T} N_{e}\thte
\frac{1}{x^{2}}\frac{\partial}
{\partial x} \bigl(x^{2} j(x)\bigr),
\label{sz_eq_phcons}
\end{equation}
where $N_{e}$ is the electron number density and the current $j(x)$ is given by
\begin{multline}
j(x) = x^2\Bigl[\Bigl(\frac{\partial n}{\partial x} + n(1+n)\Bigr)
+ \thte \Bigl[ \frac{5}{2} \Bigl(\frac{\partial n}{\partial x} + n(1+n)
\Bigr) + \frac{21}{5} x \frac{\partial}{\partial x}\Bigl(\frac{\partial n}
{\partial x} + n(1+n)\Bigr) \\
+ \frac{7}{10} x^2 \Bigl( \frac{\partial^{3} n}{\partial x^{3}} +
2\frac{\partial^{2} n}{\partial x^{2}}(1+2n) + \frac{\partial n}{\partial x}
\Bigl(1-2\frac{\partial n}{\partial x}\Bigr)
\Bigr) \Bigr]+ \ord(\thte^{2}) \Bigr].
\label{sz_eq_cur}
\end{multline}
The zero-order term in equation~\eqref{sz_eq_cur} is just that term
which usually
appears in the Kompaneets equation~\cite{komp57}. The $\ord(\thte)$ term is
the lowest-order relativistic correction to the current. The form
of equation~\eqref{sz_eq_phcons} ensures conservation of the total
number of photons, which is true for each order in $\thte$.
A similar equation has been derived independently by
Stebbins~\cite{stebbins97} using non-covariant methods, although he only
includes terms to lowest-order in $\omega/m$ which is an excellent
approximation for CMB photons.

\subsection{The Sunyaev-Zel'dovich Effect}
\label{sz_sec_sun_zel}

In this section we apply the generalised Kompaneets equation (to
first-order in relativistic corrections) to the
calculation of the Sunyaev-Zel'dovich effect in optically thin clusters.
We consider higher-order effects in the next section.

Following the standard assumptions, we assume that the optical depth
is sufficiently small that the spectral distortions are small. In this limit,
we may solve equation~\eqref{sz_eq_phcons} iteratively.
The lowest order solution is obtained by substituting the initial photon
distribution $n_{0}(x)$ into the current (eq.~\eqref{sz_eq_cur}).
The integral over time is then trivial, and
may be replaced by an integral along the line of sight through the cluster,
giving
\begin{equation}
\Delta n(x) = \frac{y}{x^{2}} \frac{\partial}{\partial x} \bigl(x^{2} j(x)
\bigr),
\label{sz_eq_z1}
\end{equation}
where $j(x)$ is evaluated with $n_{0}(x)$, and
\begin{equation}
y \equiv \sigma_{T} \int N_{e} \thte \, dl,
\label{sz_eq_z2}
\end{equation}
where the integral is taken along the line of sight through the cluster.

For the CMB we take the initial (undistorted) photon distribution to be
Planckian with temperature $T_{0}$:
\begin{equation}
n_{0}(x) = \frac{1}{\et{\alpha x} - 1},
\label{sz_eq_z3}
\end{equation}
where $\alpha \equiv T_{e}/T_{0}$ is the (large) ratio of electron temperature
to the CMB temperature. Evaluating equation~\eqref{sz_eq_z1} in the limit of
large $\alpha$, we find the following fractional distortion:
\begin{multline}
\frac{\Delta n(X)}{n(X)} = \frac{yX\et{X}}{\et{X}-1} \Bigl[
X\coth(\half X) - 4 + \thte \Bigl(
-10 + \frac{47}{2} X \coth(\half X) - \frac{42}{5} X^2 \coth^{2}(\half X)
\\
+ \frac{7}{10} X^3 \coth^{3}(\half X) +
\frac{7 X^{2}}{5 \sinh^{2}(\half X)}
\bigl(X \coth(\half X) - 3\bigr) \Bigr) \Bigl],
\label{sz_eq_z4}
\end{multline}
correct to first-order in relativistic effects, where
\begin{equation}
X \equiv \frac{\hbar\omega}{k_{B} T_{0}}.
\end{equation}
The first two terms in square brackets in equation~\eqref{sz_eq_z4} give the
usual non-relativistic Sunyaev-Zel'dovich expression, while the terms
proportional to $\thte$ are the lowest order relativistic correction.
Equation~\eqref{sz_eq_z4} agrees with the result derived by
Stebbins~\cite{stebbins97}.
In the Rayleigh-Jeans limit (small $X$), we find
\begin{equation}
\frac{\Delta n(X)}{n(X)} \simeq -2 y
\left(1- \frac{17}{10} \thte + \ord(\thte^{2})\right).
\end{equation}
%

\begin{figure}[t!]
\begin{center}
\begin{picture}(340,210)
 
 
\put(-5,235){\hbox{\epsfig{figure=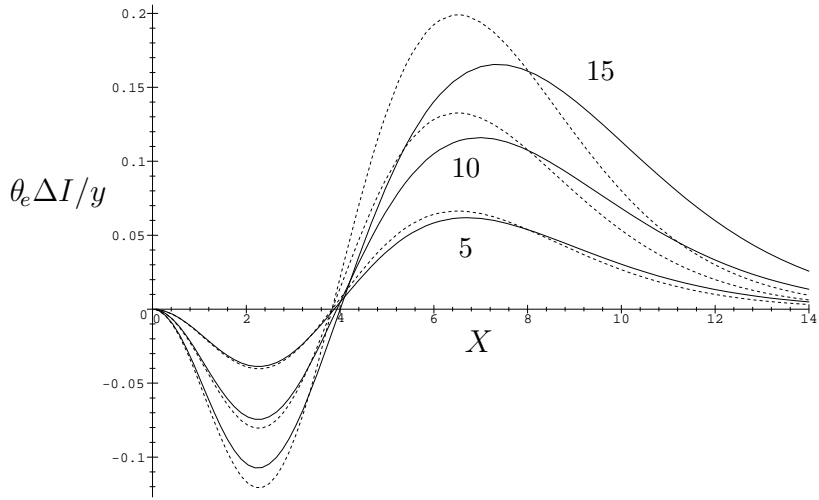,angle=-90,width=12cm}}}
\put(165,73){\makebox(0,0)[t]{\normalsize $X$}}
\put(24,125){\makebox(0,0)[r]{\normalsize $\thte\Delta I/y$}}
\put(160,108){\makebox(0,0)[t]{\small $5$}}
\put(160,138){\makebox(0,0)[t]{\small $10$}}
\put(211,175){\makebox(0,0)[t]{\small $15$}}

\end{picture}
\end{center}
\caption{The intensity change $\thte\Delta I/y$ (in units of $2(k_{B}T_{0})^{3}
/(hc)^{2}$) plotted against $X$ for three values
of $k_{B}T_{e}$ (in $\kev$).
The solid curves are calculated using the first-order
correction to the Kompaneets equation, while the dashed lines are
calculated from the usual Kompaneets expression.}
\label{sz_fig_1}
\end{figure}

In Figure~\ref{sz_fig_1} we plot
$\thte\Delta I/y$ as a function of $X$, where
\begin{equation}
\Delta I = \frac{X^{3}}{\et{X}-1} \frac{\Delta n}{n},
\end{equation}
is the change in spectral intensity in units of $2(k_{B}T_{0})^{3}/(hc)^{2}$.
Also plotted in Figure~\ref{sz_fig_1} are the non-relativistic predictions
made with the standard Kompaneets equation. 
The curves in Figure~\ref{sz_fig_1} are for $k_{B}T_{e}=5$, $10$, and $15\kev$,
which are the same as the parameters used
by Rephaeli~\cite{reph95} in his Figure 1. His calculations, which
were based on the multiple scattering formalism~\cite{wright79}
and required a numerical analysis, give results in excellent agreement
with ours, which only require the use of the simple expression~\eqref{sz_eq_4}.
This suggests that there is no difficulty in principle with applying
the Boltzmann
equation to the problem of Comptonization in clusters even though the
optical depth may be very small. Similar conclusions were reached by
Fabbri~\cite{fabbri81}, but his demonstration was restricted to low
temperature clusters where relativistic effects are not important.

It is clear from Figure~\ref{sz_fig_1} that for $X \simlt 8$, the relativistic
corrections lead to a reduction in the magnitude of the intensity change,
compared to the non-relativistic prediction.
The Hubble constant is inferred from combined Sunyaev-Zel'dovich and
X-ray data by a relation of the form~\cite{las97}
\begin{equation}
H_{0} \propto \Delta I^{-2},
\label{sz_new_eq1}
\end{equation}
where $\Delta I$ is the observed intensity change. The reduction in the
magnitude of $\Delta I$ in the Rayleigh-Jeans region,
for given cluster parameters, amounts to a decrease in the constant of
proportionality in~\eqref{sz_new_eq1} and hence a reduction
in the value of the Hubble constant that should be inferred.

\subsection{Higher-order Effects}

We have found that for $k_{B}T_{e} \simgt 10 \kev$ the second-order
relativistic effects make a significant contribution to the
spectral distortion, while third-order effects are only significant for
$k_{B}T_{e} \simgt 15 \kev$.

These calculations require a straightforward extension of the method
of Section~\ref{sz_sec_komp} to include terms at $\ord(\thte^{3})$ (for
second-order relativistic effects). For the calculation to $\ord(\thte^{3})$,
it is necessary to retain the first six
terms of the series~\eqref{sz_eq_3}, and to calculate $I_{1}$ through $I_{6}$
to $\ord(\thte^{3})$. The first iteration
of equation~\eqref{sz_eq_phcons} for $T_{e} \gg T_{0}$
gives the following next order (in $\thte$) correction to $\Delta n/n$:
\begin{multline}
\left(\frac{\Delta n(X)}{n(X)}\right)^{\!(2)} =
\thte^{2} \frac{yX\et{X}}{\et{X}-1} \Bigl[-\frac{15}{2} + \frac{1023}{8}
X\coth(\half X) - \frac{868}{5} X^{2} \coth^{2}(\half X) 
\\
+ \frac{329}{5}
X^{3}\coth^{3}(\half X) - \frac{44}{5} X^{4} \coth^{4}(\half X) +
\frac{11}{30} X^{5} \coth^{5}(\half X)
\\
+ \frac{X^{2}}{30 \sinh^{2}(\half X)}
\bigl( -2604 + 3948 X\coth(\half X) 
- 1452 X^{2} \coth^{2}(\half X) + 143
X^{3} \coth^{3}(\half X) \bigr) 
\\
+ \frac{X^{4}}{60\sinh^{4}(\half X)}
\bigl( -528 + 187 X\coth(\half X) \bigr) \Bigl].
\label{sz_eq_s_1}
\end{multline}
In the Rayleigh-Jeans limit, we find
\begin{equation}
\frac{\Delta n(X)}{n(X)} \simeq -2y \left( 1- \frac{17}{10} \thte +
\frac{123}{40} \thte^{2} + \ord(\thte^{3}) \right).
\label{sz_eq_s_2}
\end{equation}
%


\begin{figure}[t!]
\begin{center}
\begin{picture}(340,210)
 
 
\put(-5,235){\hbox{\epsfig{figure=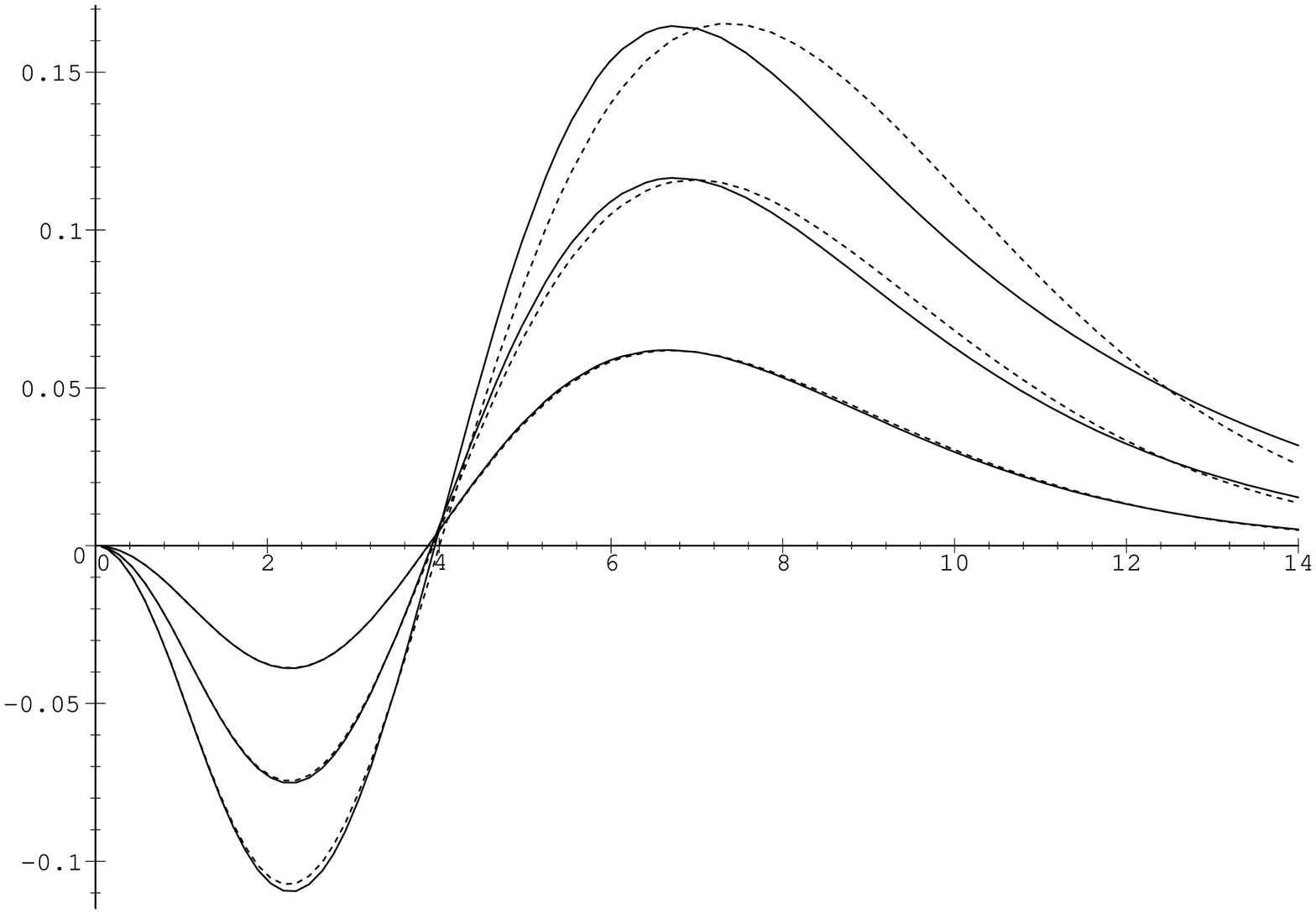,angle=-90,width=12cm}}}
\put(168,80){\makebox(0,0)[t]{\normalsize $X$}}
\put(24,135){\makebox(0,0)[r]{\normalsize $\thte\Delta I/y$}}
\put(160,119){\makebox(0,0)[t]{\small $5$}}
\put(160,150){\makebox(0,0)[t]{\small $10$}}
\put(160,180){\makebox(0,0)[t]{\small $15$}}
  
\end{picture}
\end{center}
\caption{The intensity change $\thte\Delta I/y$ (in units of $2(k_{B}T_{0})^{3}
/(hc)^{2}$) plotted against $X$ for three values
of $k_{B}T_{e}$ (in $\kev$).
The solid curves are calculated using the second-order
correction to the Kompaneets equation, while the dashed lines are
calculated from the first-order correction.}
\label{sz_fig_3}
\end{figure}


In Figure~\ref{sz_fig_3} we compare the spectrum of $\Delta I$ calculated
with equation~\eqref{sz_eq_z4} to the spectrum with the
correction~\eqref{sz_eq_s_1} included, for $k_{B}T_{e} = 5$, $10$ and
$15\kev$ ($\thte \approx 0.01$,
$0.02$ and $0.03$ respectively). In each case, the
second-order relativistic effects are not significant in the Rayleigh-Jeans
part of the spectrum. This is to be expected from inspection
of equation~\eqref{sz_eq_s_2}, where the $\thte^{2}$ term is
clearly insignificant for the values of $\thte$ considered.
For $k_{B}T_{e}=5\kev$, the second-order effects are
insignificant over the entire spectrum. However, for $k_{B}T_{e} \simgt
10\kev$, the second-order effects make a significant contribution to
the relativistic correction to the Kompaneets-based prediction outside the
Rayleigh-Jeans region. We have verified that the
third-order corrections are negligible over the entire spectrum for
$k_{B}T_{e} \simeq 10\kev$. This has been confirmed independently by
a direct Monte-Carlo evaluation of the Boltzmann collision integral
by Gull and Garrett~\cite{gull97}.
The
second-order effects should be included in the analysis of high frequency data
for hot clusters. The magnitude of the second-order
correction to the Sunyaev-Zel'dovich result for the rather mild
values of $\thte$ considered here, is symptomatic of the asymptotic nature
of the series expansion of
$\partial n/\partial t$ in $\thte$.
However, for the majority of clusters
considered in Sunyaev-Zel'dovich analyses, the inclusion of the first
two relativistic corrections is sufficient, particularly for
experiments working in the Rayleigh-Jeans region of the spectrum.

\subsection{The Crossover Frequency}


\begin{figure}[t!]
\begin{center}
\begin{picture}(340,210)
 
 
\put(-5,245){\hbox{\epsfig{figure=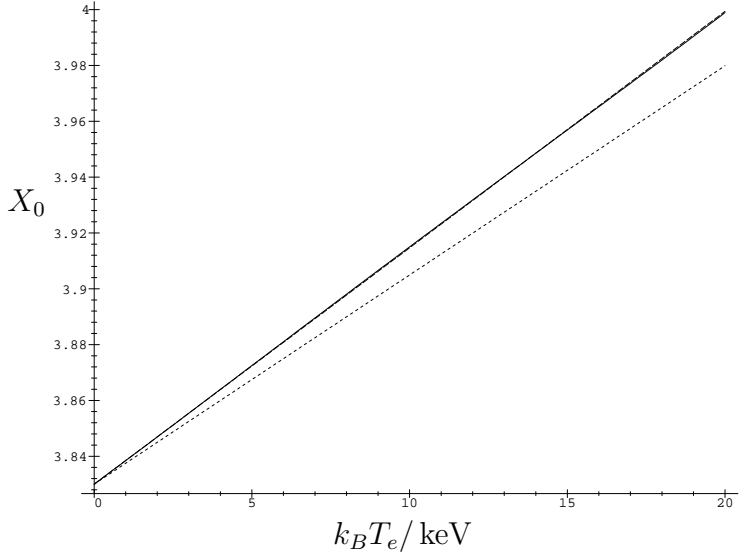,angle=-90,width=12cm}}}
\put(163,10){\makebox(0,0)[t]{\normalsize $k_{B}T_{e}/\kev$}}
\put(28,130){\makebox(0,0)[r]{\normalsize $X_{0}$}}

\end{picture}
\end{center}
\caption{The crossover frequency $X_{0}$ plotted against
$k_{B}T_{e}$.
The solid line is calculated with the inclusion of third-order corrections
to the Kompaneets equation. The upper dotted line is a linear fit to the solid
line with $X_{0}=3.83(1+1.13\theta_{e})$, while the lower dotted line is the
linear fit given by Rephaeli: $X_{0} = 3.83(1+\theta_{e})$.}
\label{sz_fig_4}
\end{figure}


The accurate determination of the crossover frequency $X_{0}$
(where the thermal component of the spectral distortion vanishes)
is essential for reliable
subtraction of the kinematic contribution to the Sunyaev-Zel'dovich
effect~\cite{reph95}.
In Figure~\ref{sz_fig_4} we plot the crossover frequency as a function of
$k_{B}T_{e}$, with the first three relativistic corrections included.
For $k_{B}T_{e} \simlt 20\kev$ we find that $X_{0}$ is well approximated by
the linear relation
\begin{equation}
X_{0} \simeq 3.83(1+1.13\thte).
\label{sz_eq_s_3}
\end{equation}
For comparison, Rephaeli~\cite{reph95},
found $X_{0}$ to be approximated by $X_{0}\simeq 3.83(1+\thte)$
in the interval $k_{B}T_{e} = 1 \mbox{--} 50\kev$, while Fabbri~\cite{fabbri81}
found $X_{0}\simeq 3.83(1+1.1\thte)$ for $k_{B}T_{e} \simlt 150\kev$.
It is clear that our calculation favours Fabbri's expression.
For $k_{B}T_{e} \simgt 20\kev$, $X_{0}$ calculated with the first three
relativistic corrections departs from the linear prediction~\eqref{sz_eq_s_3}.
However, we do not regard this as indicative of a breakdown of the linear
approximation, since it is clear from Figure~\ref{sz_fig_3}
that the inclusion of higher-order terms may have a significant effect on the
value of the crossover frequency.

\subsection{Conclusion}

The application of the relativistic extension of the Kompaneets
equation, presented here, to the Sunyaev-Zel'dovich effect in optically thin
clusters results in simple analytic expressions
for the relativistic corrections to the usual Kompaneets based
expression for the spectral distortion. These expressions are in
excellent agreement with the numerical calculations of
Rephaeli~\cite{reph95} for electron temperatures $\simlt 10\kev$,
providing further evidence that the low optical depth of clusters does not
forbid the application of the Boltzmann equation to the calculation of the
Sunyaev-Zel'dovich effect~\cite{fabbri81}.
The asymptotic nature of the series
expansion of $\partial n/\partial t$ in $\thte$ requires the inclusion
of higher-order corrections to calculate the effect in hotter clusters
in the Wien region of the spectrum.
While the calculation of higher-order corrections is not problematic,
the bad convergence properties of the series means that ultimately one must
resort to a numerical calculation of the
collision integral~\cite{gull97,corman70} (or employ the multiple scattering
formalism~\cite{wright79}) to calculate the distortion in very hot clusters.

Our calculations fully support the conclusions reached in
Rephaeli~\cite{reph97}, that
including relativistic effects leads to a small decrease in the value of the
Hubble constant, inferred from combined X-ray and Sunyaev-Zel'dovich
information. For a cluster temperature of $\simeq 8\kev$, the reduction in
$H_{0}$ due to relativistic effects is $\simeq 5$ percent for measurements made
in the Rayleigh-Jeans region.

\section{Covariant and Gauge-Invariant Calculations of CMB Anisotropies}
\label{sec_kin}

In this section we consider a covariant, kinetic theory approach for the
gauge-invariant calculation of CMB anisotropies. We employ the
elegant gauge-invariant perturbation methods of
Ellis~\etal~\cite{ellis89a,ellis89b} (see also the recent Erice lectures
by Ellis~\cite{ellis94er,ellis-er96} and the contribution to this volume from
Ellis and Dunsby), although we prefer to express this within
the gauge-theoretic approach
to gravity (GTG) developed recently in Cambridge, instead of conventional
general relativity. GTG has been discussed in previous Erice
lectures~\cite{DGL-erice,DGL96-erice}, although the most detailed
discussion to date is contained in Lasenby, Doran and Gull~\cite{DGL-grav}.
In GTG, gravity is introduced via gauge-fields over a flat (background)
spacetime.
The gauge-theoretic purpose of the gauge-fields is to ensure
covariance of the theory under smooth remappings of physical events
onto vectors in the background spacetime, and under local rotations of
geometric quantities at a point in the vector space. Physical observables
must be extracted from the theory in a gauge-invariant manner, which ensures
that the background vector space does not play an active role in the theory.
GTG differs from general relativity only in certain circumstances, such as
the treatment of horizons~\cite{DGL-grav}, solutions of general relativity
with non-trivial topology, and the effects of quantum
spin~\cite{DGL-spintor,DGL-selfcons}. In situations where none of these
effects are present, such as the discussion of small departures from
an exact Friedmann-Robertson-Walker (FRW) universe we consider here,
the predictions of GTG coincide with general relativity. Formulating
GTG in the powerful language of geometric algebra~\cite{hes-gc} allows GTG
to be expressed in a coordinate-free manner, where all algebraic manipulations
may be performed efficiently in the geometric algebra of
spacetime~\cite{hes-sta} (the STA). Our notation and conventions for GTG are
the same as in Lasenby, Doran and Gull~\cite{DGL-grav} (see also the previous
Erice lectures~\cite{DGL-erice,DGL96-erice}).

The covariant and gauge-invariant treatment of cosmological perturbations
pioneered by Ellis~\etal\ has
many advantages over other approaches. The variables are covariantly-defined
and relate to observable quantities in a simple manner. Furthermore, the
variables are gauge-invariant, in that they are independent of the way that
the `lumpy' universe is mapped onto the background FRW universe (equivalently,
a variable is gauge-invariant in this context if it transforms covariantly
under gauge transformations of the lumpy universe, with the gauge for the
background model held fixed). Note that in the covariant approach, the
variables are truly gauge-invariant, which should be contrasted to the
variables in Bardeen's~\cite{bardeen80} `gauge-invariant' approach, which are
gauge-invariant only if the scalar, vector or tensor nature of the
perturbations is not altered by the remapping. A further advantage of the
covariant approach is that the perturbation equations may be formulated
without the need for a scalar, vector or tensor splitting of the perturbations,
or the associated harmonic analysis, whereas these are required at the outset
in the Bardeen approach. The equations pertaining to a given type of
perturbation may be obtained readily from the full perturbation equations
by placing appropriate restrictions on the gauge-invariant variables.

The covariant calculation of large scale CMB anisotropies is dealt with
in the lecture by Ellis~\cite{ellis-er97}, where
a two fluid model of the matter and radiation are used, and a sharp last
scattering surface is assumed. On small scales, the particle-like nature
of the radiation (and neutrinos) becomes important, as well as the
complicated local physics of recombination and decoupling.
An accurate treatment of small scale effects requires a phase-space
description of the photons, and the solution of the
Boltzmann equation describing the evolution of the distribution function.
Similarly, a phase-space description of the neutrinos is necessary to
deal properly with the heat flux and anisotropic stress terms in the
neutrino stress-energy tensor.
In this section, we use covariant distribution functions which we decompose
covariantly into angular moments, as discussed by
Ellis~\etal~\cite{ellis94er,stoeger95b,maartens95,ellis83}.
In these papers, the discussion is limited to the
decomposition of the Liouville equation, appropriate for freely propagating
radiation. Here we present the covariant Boltzmann equation (for Thomson
scattering of radiation off free electrons), and its angular moments.
At present, polarisation is not
included for simplicity. Since the generation of photon polarisation
is sourced by anisotropy for Thompson scattering, the error resulting from the
neglect of polarisation is small~\cite{hu95}.
Using this formalism, we discuss the covariant equations describing the
evolution of perturbations in a Cold Dark Matter (CDM) model. Our (linearised)
equations are independent of scalar, vector or tensor decomposition, and
of the spatial curvature of the background FRW model. Specialising to
scalar perturbations in a $K=0$ universe, we obtain a set of scalar mode
equations for gauge-invariant variables. These equations must be solved
numerically to obtain the present day anisotropies for given initial
conditions. This work will be presented elsewhere~\cite{chall97-cmb}.
We end this section with a discussion of large scale anisotropies, which is
complementary to the discussion of Ellis and Dunsby~\cite{ellis-er97}.
The large scale anisotropy is obtained by integrating the covariant Boltzmann
equation along null geodesics, and assuming instantaneous recombination.
With the tight-coupling approximation, we obtain the same expression as
that obtained by Ellis and Dunsby with a two-fluid model.

\subsection{The Gauge-Invariant Variables}
\label{sec_var}

The starting point in the covariant description of cosmological perturbations
is the choice of a velocity $u$. This should be chosen in a physical manner
in such a way that if the universe is exactly FRW, the velocity reduces to
that of the fundamental observers. Having made a choice for $u$, we define
a projection tensor $H(a)$ which projects vectors into the space perpendicular
to $u$:
\begin{equation}
H(a) \equiv a - a\dt u u.
\label{eq_proj}
\end{equation}
For details of our geometric algebra notation see
Lasenby~\etal~\cite{DGL-erice}--\cite{DGL-grav}.
Expressing the theory in geometric algebra provides frame-free and
coordinate-free versions of all variables and operations.
Using the projection~\eqref{eq_proj},
we form a directional `spatial' derivative,
$a\dt\cldh$ from the usual directional (covariant) derivative $a\dt\cld$ via
\begin{equation}
a\dt\cldh M \equiv H[H(a)\dt \cld M],
\end{equation}
where $M$ is an arbitrary multivector. We also define a derivative
$a\dt\cldb$ which acts on tensors $T(b,\dots,c)$ according to
\begin{equation}
a\dt\cldb T(b,\dots,c) \equiv H[H(a) \dt \dot{\cld} \dot{T}(H(b),\dots,
H(c))],
\end{equation}
where $a\dt\dot{\cld} \dot{T}(b,\dots,c) = a\dt\cld T(b,\dots,c)
- T(a\dt\cld b,\dots,c) - T(b,\dots,a\dt\cld c)$. In an exact
FRW universe, spatial gradients of any field $M$ necessarily vanish
by homogeneity, so that $a\dt\cldh M = 0$. It follows
that the spatial gradient of the density $\cldh \rho$ is a gauge-invariant
variable. Quantities such as $\cldh \rho$ (or, more correctly, the
dimensionless quantity $\cldh \rho /(\rho H)$, where $H$ is a suitable
average of the local Hubble parameter) will be small compared to unity if the
universe is sufficiently close to an FRW universe. We shall refer to such
variables as being first-order, or $\ord(1)$. In the linear theory used here,
products of first-order quantities may be neglected in an expression, compared
to any first-order (or zero-order) quantities. It is convenient to introduce
a scale factor $S$, satisfying
\begin{equation}
u\dt\cld S = H S, \qquad \cldh S = \ord(1),
\end{equation}
so that we may remove the variation in $\cldh \rho$ along an integral curve
of $u$ (a `worldline') by defining the variable
\begin{equation}
\clx \equiv \frac{S \cldh \rho}{\rho}.
\end{equation}
The first-order, gauge-invariant variable $\clx$ is an example of a
fractional, comoving spatial gradient.

The covariant derivative of $u$ decomposes as
\begin{equation}
a\dt\cld u = \half a\dt \varpi + \sigma(a) + \third H(a) \theta + a\dt u w,
\end{equation}
where the vector $w \equiv u\dt \cld u$ is the acceleration, the bivector
$\varpi \equiv \cld \wdg u + w\wdg u$ is the vorticity, measuring the local
rotation of the worldlines, and the scalar $\theta \equiv \cld \dt u = 3H$
measures the volume expansion rate. The traceless symmetric tensor
$\sigma(a)$ measures the shear of the worldlines, and is orthogonal
to $u$: $\sigma(u) = 0$. We shall refer to tensors orthogonal to $u$ as
being spatial tensors. The vorticity, acceleration and shear are all
first-order gauge-invariant variables (they vanish in an FRW universe).
However, like $\rho$, the volume expansion rate $\theta$ is zero-order, but
we obtain a first-order gauge-invariant variable $\clz$ by taking the spatial
gradient, $\clz \equiv S \cldh \theta$.

The matter stress-energy tensor $\clt(a)$ decomposes as
\begin{equation}
\clt(a) = \rho a\dt u u + q a\dt u + u q\dt a - p H(a) + \pi(a),
\end{equation}
where the scalar $\rho\equiv \clt(u)\dt u$ is the density of matter
(measured by a comoving observer), the vector $q \equiv H[\clt(u)]$ is the
energy (or heat) flux, the scalar $p\equiv - H(\da)\dt \clt(a) /3$
is the isotropic pressure, and the traceless, symmetric, spatial tensor
$\pi(a) \equiv H[\clt H(a)] + p H(a)$ is the anisotropic stress.
In an exact FRW universe, the stress-energy tensor is forced to take the
ideal fluid form, hence $q$ and $\pi(a)$ are first-order variables.
A first-order, gauge-invariant variable may be derived from the pressure by
taking a spatial gradient.

The final gauge-invariant variables we shall make use of derive from the
Weyl tensor $\clw(B)$, which is a bivector-valued linear function of a
bivector $B$. It is convenient to split the Weyl tensor into electric
and magnetic parts, in analogy with the split of the Faraday bivector in
electromagnetism. We define the electric part by
\begin{equation}
\cle(a) \equiv - u \dt \clw(a\wdg u),
\label{kin_eq_1}
\end{equation}
and the magnetic part by
\begin{equation}
\clb(a) \equiv - i u\wdg \clw(a\wdg u),
\label{kin_eq_2}
\end{equation}
where $i$ is the unit pseudoscalar for spacetime (which performs the
duality operation in the STA). Both $\cle(a)$ and $\clb(a)$ are traceless,
symmetric spatial tensors. The definitions~\eqref{kin_eq_1}
and~\eqref{kin_eq_2} combine together to give the neat relation
\begin{equation}
\clw(a\wdg u) u = \cle(a) + i \clb(a).
\end{equation}
Since the Weyl tensor vanishes in an FRW universe, the electric and magnetic
parts of the Weyl tensor are first-order gauge-invariant variables.
The equations of motion for the gauge-invariant variables may be derived from
the Ricci identity, the Bianchi identity, and the contracted Bianchi
identity. These equations are the GTG equivalents of the equations
given by Ellis~\cite{ellis-er97} in his lecture, and will be given
in linearised from in Section~\ref{sec_gen}.
The equations split into propagation
equations (those involving the derivative along a worldline $u\dt\cld$), and
constraint equations which involve only spatial derivatives. The constraint
equations restrict the specification of initial data on a hypersurface (the
data must satisfy the constraints), while the propagation equations
allow one to propagate the data off the surface. 

\subsection{The Boltzmann Equation}

The photon distribution function is a scalar valued function of position $x$
and covariant momentum $p$, denoted by $\fgam(x,p)$. Physically, an
observer sees $\fgam(x,p) dV |\dtri p|$ photons in a volume $dV$ and
a 3-momentum volume $|\dtri p|$. It is convenient to split the
photon momentum $p$ as
\begin{equation}
p = E(U + e),
\end{equation}
where $E\equiv p\dt u$ is the photon energy and $e$ is a unit spacelike
vector $e^{2}=-1$ orthogonal to $u$. The photon distribution function may then
be written as $\fgam(x,E,e)$. We will usually leave the $x$ dependence
implicit, writing $\fgam(E,e)$. The stress-energy tensor for the photons may
be written as
\begin{equation}
\clt_{\gamma}(a) = \int dE d\Omega\, E \fgam(E,e) p \, p\dt a ,
\end{equation}
where $d\Omega$ denotes an integral over solid angles.

Following Ellis~\etal~\cite{ellis83}, we describe the angular dependence
of $\fgam(E,e)$ by expanding in covariant, scalar-valued tensors
$\fgaml{l}(a_{1},a_{2},\dots,a_{l})$:
\begin{equation}
\fgam(E,e) = \sum_{l=0}^{\infty} \fgaml{l}(e,e,\dots,e).
\label{be_eq_4}
\end{equation}
The tensors $\fgaml{l}(a_{1},a_{2},\dots,a_{l})$ have an (implicit) dependence
on $x$ and $E$ and satisfy the properties:
\begin{align}
\fgaml{l}(a_{1},\dots,a_{j},\dots,a_{k},\dots,a_{l}) & =
\fgaml{l}(a_{1},\dots,a_{k},\dots,a_{j},\dots,a_{l}) \notag\\
\fgaml{l}(u,a_{2},\dots,a_{l}) & = 0 \\
\fgaml{l}(\db,b,a_{3},\dots,a_{l}) & = 0 . \notag
\end{align}
The action of the Liouville operator $\cll$ on
$\fgaml{l}(e,\dots,e)$, evaluates to
\begin{equation}
\cll \fgaml{l}(e,\dots,e) 
=  \partial_{E}\fgaml{l}(e,\dots,e)\partial_{\lambda} E
+ p\dt\dot{\cld}\fgamldt{l}(e,\dots,e) + l\fgaml{l}(p\dt\cld e,e,\dots,e),
\label{be_eq_8}
\end{equation}
where $\lambda$ is the affine parameter to the photon path ($p=\hu^{-1}
(\partial_{\lambda} x)$ with $\ho(a)$ the GTG position-gauge field).
It follows that the action of the Liouville
operator on $\fgam(E,e)$ may be written in the form
\begin{equation}
\cll\fgam(E,e)=\sum_{l=0}^{\infty} [ \partial_{E}\fgaml{l}(e,\dots,e)
\partial_{\lambda} E + p\dt\dot{\cld}\fgamldt{l}(e,\dots,e) +
l\fgaml{l}(p\dt\cld e,e,\dots,e)].
\label{be_eq_9}
\end{equation}
For $l>0$ the tensors $\fgaml{l}(a_{1},\dots,a_{l})$ are first-order
quantities. However, to zero-order, we have
\begin{equation}
p\dt\cld e = \third \theta E u,
\end{equation}
so that the final term in~\eqref{be_eq_9} only contributes at second-order,
and may be ignored in the linearised calculation considered here.

Since neutrinos are essentially collisionless over the period of interest,
their distribution function $\fnu(x,p)$ evolves according to the Liouville
equation $\cll \fnu(x,p)=0$. The photons, however, are in close contact with
the baryons through Thomson scattering off free electrons (which are
at rest in the frame of the baryons because of Coulomb interactions).
In the presence of collisions, the photon distribution function evolves
according to the Boltzmann equation
\begin{equation}
\cll \fgam(x,p) = C,
\end{equation}
where the collision term approximates to
\begin{equation}
C= n_{e} \sigma_{T} p\dt u_{b} (f_{+}(x,p) - f(x,p)),
\end{equation}
with $u_{b}$ denoting the baryon velocity, $\sigma_{T}$ the Thomson cross
section, and $f_{+}(x,p)$ describes scattering into the phase space element
under consideration. Evaluating $f_{+}(x,p)$ to first-order, multiplying by
$E^{2}$ and integrating over energy, we find~\cite{chall97-cmb}
\begin{multline}
\int_{0}^{\infty} dE \, E^{2} \cll \fgam(E,e) = \frac{3}{16\pi}
\sigma_{T} n_{e} 
\left(\frac{4}{3} (1-4e\dt \velb) \rho_{\gamma} + e\dt
\pi_{\gamma}(e)\right)
\\
- \sigma_{T} n_{e} \int_{0}^{\infty} dE \, E^{3} \fgam(E,e),
\label{be_eq_26}
\end{multline}
where $v_{b} \equiv u_{b} - u$ is the first-order relative velocity
of the baryons.

Equation~\eqref{be_eq_26} may itself be expanded in
symmetric, trace-free tensors orthogonal to $u$
(like $\fgaml{l}(a_{1},\dots,a_{l})$).
The $l=0$, $1$ and $2$ terms require a little care in the
linearised calculation since $\fgaml{0}$ is a zero-order quantity.
The result is~\cite{chall97-cmb}:
\newline
for $l=0$ 
\begin{equation}
u\dt\cld\rho_{\gamma} + \tfrac{4}{3} \theta \rho_{\gamma} + \cldh\dt
q_{\gamma} = 0,
\label{be_eq_27}
\end{equation}
for $l=1$
\begin{equation}
u\dt\cld q_{\gamma} + \tfrac{4}{3} \theta q_{\gamma} + \pi_{\gamma}(\cldb)
+ \tfrac{4}{3} w \rho_{\gamma} - \tfrac{1}{3} \cldh \rho_{\gamma} =
\sigma_{T} n_{e} \left(\tfrac{4}{3} \rho_{\gamma} \velb - q_{\gamma} \right),
\label{be_eq_28}
\end{equation}
for $l=2$
\begin{multline}
a_{1} \dt (u\dt\dot{\cld}\dot{\pi}_{\gamma}(a_{2}) )
+ \tfrac{4}{3} \theta a_{1}\dt \pi_{\gamma}(a_{2}) +
\jgaml{3}(\cldb,a_{1},a_{2}) -
\tfrac{1}{5} \bigl(
a_{1}\dt(a_{2}\dt\cldh q_{\gamma}) + a_{2}\dt(a_{1}\dt\cldh q_{\gamma})
\\
- \tfrac{2}{3} H(a_{1})\dt H(a_{2}) \cldh\dt q_{\gamma} \bigr)
- \tfrac{8}{15} a_{1}\dt \sigma(a_{2}) \rho_{\gamma} =
-\tfrac{9}{10} \sigma_{T} n_{e} a_{1}\dt \pi_{\gamma}(a_{2}),
\label{be_eq_29}
\end{multline}
and for $l\geq 3$
\begin{multline}
u\dt\dot{\cld} \jgamldt{l}(a_{1},\dots,a_{l}) + \tfrac{4}{3}\theta
\jgaml{l}(a_{1},\dots,a_{l}) + \jgaml{l+1}(\cldb,a_{1},\dots,a_{l})
\\
- \tfrac{l}{(2l+1)l!} \bigl( a_{1}\dt\cldb \jgaml{l-1}(a_{2},\dots,a_{l}) -
\tfrac{(l-1)}{(2l-1)} \jgaml{l-1}(\cldb,a_{1},\dots,a_{l-2}) H(a_{l-1})
\dt H(a_{l}) + \mbox{perms} \bigr)
\\
=  -\sigma_{T}n_{e} \jgaml{l}(a_{1},\dots,
a_{l}).
\label{be_eq_30}
\end{multline}
The scalar-valued, traceless, symmetric, spatial tensor
$\jgaml{l}(a_{1},\dots,a_{l})$ is defined by
\begin{equation}
\jgaml{l}(a_{1},\dots,a_{l}) \equiv \frac{4\pi (-2)^{l} (l!)^{2}}{(2l+1) (2l)!}
\int_{0}^{\infty} dE \, E^{3} \fgaml{l}(a_{1},\dots,a_{l}).
\label{be_eq_13}
\end{equation}
With this definition, we have
\begin{align}
\rhogam & = \jgaml{0} \\
q_{\gamma} & = \da \jgaml{1}(a) \\
\pi_{\gamma}(a)&=\db \jgaml{2}(b,a).
\end{align}
It is not hard to show that the combination
\begin{equation}
a_{1}\dt\cldb \jgaml{l-1}(a_{2},\dots,a_{l}) -
\tfrac{(l-1)}{(2l-1)} \jgaml{l-1}(\cldb,a_{1},\dots,a_{l-2}) H(a_{l-1})
\dt H(a_{l}) + \mbox{perms},
\label{be_eq_15}
\end{equation}
is a trace-free, symmetric tensor, orthogonal to $u$, as required.
Note that the linearised equations show
a coupling between the $l-1$, $l$ and $l+1$ terms, whereas the exact equations
also show coupling between the $l-2$ and $l+2$ terms. The terms involving
$\jgaml{l-2}(a_{1},\dots,a_{l-2})$ and $\jgaml{l+2}(a_{1},\dots,a_{l+2})$
also involve the shear tensor $\sigma(a)$. Ellis has remarked~\cite{ellis-er96}
that the exact result that if the hierarchy of $\jgaml{l}$ tensors truncates
after a finite number of terms then the shear must vanish, is missed in the
linearised approach. This is certainly true, however it is not problematic
for the calculation of CMB anisotropies, since it is never claimed that the
expansion of the photon distribution function truncates. Instead, the series
is truncated (with suitable care to avoid reflection of power) after a
finite number of terms for numerical convenience (see,
for example, Ma and Bertshinger~\cite{ma95}). The truncation is chosen
to be high enough up the series that it has negligible effect on the
$\jgaml{l}$ for the range of $l$ of interest.

\subsection{Baryons and Dark Matter}

Over the epoch of interest here, the baryons and electrons are
non-relativistic, and we assume that they may be described as an ideal fluid,
with energy density $\rhob$ (measured in the rest frame of the baryons),
pressure $\prb$, and covariant-velocity $u+\velb$, where $\velb$ is a
first-order
quantity. The linearised baryonic stress-energy tensor evaluates to
\begin{equation}
\clt_{b}(a) = \rhob a\dt u u - \prb H(a) + (\rhob + \prb)(a\dt u \velb +
a\dt\velb u),
\label{bar_eq_1}
\end{equation}
which shows that there is a heat-flux
$(\rhob + \prb)\velb$ measured by an observer moving
with velocity $u$. To find the equations of motion for $\rhob$ and $\velb$,
we make use of the fact that the baryons (including electrons) and photons
interact non-gravitationally only with themselves, so that
\begin{equation}
\dot{\clt}_{\gamma}(\dot{\cld}) + \dot{\clt}_{b}(\dot{\cld}) = 0.
\label{bar_eq_2}
\end{equation}
Using the moment equations of the previous section, we find a propagation
equation for $\rhob$:
\begin{equation}
u\dt\cld\rhob + (\rhob + \prb)\theta + (\rhob+\prb)\cldh\dt \velb = 0,
\label{bar_eq_3}
\end{equation}
and a propagation equation for $\velb$:
\begin{equation}
(\rhob + \prb) (u\dt\cld \velb + w) + \tfrac{1}{3}(\rhob+\prb) \theta \velb +
u\dt\cld \prb - \cldh \prb + \sigma_{T}n_{e} \left(\tfrac{4}{3} \rhogam
\velb - q_{\gamma} \right) = 0,
\label{bar_eq_4}
\end{equation}
which must be supplemented by an equation of state linking
$\prb$ and $\rhob$.
Note that equation~\eqref{bar_eq_3} implies that there
is no energy exchange between the radiation and the baryon plasma in the
frame in which the baryon plasma is at rest, while
equation~\eqref{bar_eq_4} shows that there is momentum exchange due
to dipole anisotropy of the CMB in the frame of the plasma. This
behaviour is to be expected since we have assumed that the baryon-photon
interaction approximates to Thomson scattering in the baryon rest frame,
for which there will be no energy transfer, but there will be
momentum transfer if the photon distribution function is not isotropic.

We will only consider cold dark matter (CDM) here, which may be described
as a pressureless ideal fluid. Hot dark matter (HDM)
would require a distribution function description which for massive particles
greatly increases the numerical complexity of the problem. (Both CDM and
HDM are considered, for example, in the gauge-dependent treatment
in Ma and Bertshinger~\cite{ma95}).
The CDM has energy density $\rhocdm$ in its rest frame, which
has covariant velocity $u+\velcdm$, with $\velcdm$ a first-order quantity.
The CDM interacts with other species through gravity alone, so the equations
of motion are
\begin{align}
u\dt\cld \rhocdm + \rhocdm \theta + \rhocdm \cldh\dt\velcdm & = 0
\label{bar_eq_5} \\
u\dt\cld \velcdm + \tfrac{1}{3} \theta \velcdm + w & = 0. \label{bar_eq_6}
\end{align}
It is very convenient to use the CDM velocity to define the fundamental
velocity $u$ ($\velcdm = 0$). Since the CDM is pressureless, it moves
geodesically so that the acceleration $w$ will vanish for this choice of $u$.
The discussion in the next two sections assumes that this
choice has been made.

\subsection{General Equations for the Evolution of CMB Anisotropies}
\label{sec_gen}

As Ellis and Bruni~\cite{ellis89a} have shown,
the natural variables with which to
describe cosmological perturbations are the comoving fractional spatial
gradients of the energy densities, $\clx_{i}$ where $i$ labels the particle
species. The equations of motion for these variables follow from taking
the spatial gradients of the evolution equations for the $\rho_{i}$.
This procedure yields the linearised equations:
\begin{align}
u\dt\cld\clxnu & = -\frac{4}{3} \clz - \frac{S}{\rhonu} \cldh \cldh\dt q_{\nu}
\label{gen_eq_1} \\
u\dt\cld\clxgam & = -\frac{4}{3} \clz - \frac{S}{\rhogam} \cldh \cldh\dt
q_{\gamma} \label{gen_eq_2} \\
u\dt\cld\clxcdm & = - \clz \label{gen_eq_3} \\
u\dt\cld\clxb & = - \left(1+ \frac{\prb}{\rhob}\right)(\clz + S \cldh \cldh\dt
\velb) - S\theta \frac{\cldh \prb}{\rhob} + \theta \clxb \frac{\prb}{\rhob}.
\label{gen_eq_4}
\end{align}
An equation of motion for $\clz$ may be
obtained by taking the spatial gradient of the propagation equation for
$\theta$ (the Raychaudhuri equation, which follows from the Ricci identity)
to obtain
\begin{equation}
u\dt\cld \clz = -\tfrac{2}{3} \theta \clz - \kappa \left(
\rhogam \clxgam + \rhonu\clxnu + \tfrac{1}{2}\rhocdm \clxcdm + \tfrac{1}{2}
\rhob \clxb + \tfrac{3}{2} S \cldh \prb\right).
\label{gen_eq_5}
\end{equation}
Note that these equations do not close due to the presence of $\velb$,
$q_{\gamma}$ and $q_{\nu}$. These equations, along with the kinetic theory
equations given in the previous section and the equations for the
gauge-invariant variables, discussed in Section~\ref{sec_var}, form a complete
description of the evolution of CMB anisotropies. For convenience, we group
these equations together in this section.

The baryon peculiar velocity $\velb$ evolves according to
\begin{equation}
(\rhob + \prb) u\dt\cld \velb + \tfrac{1}{3}(\rhob+\prb) \theta \velb +
\velb u\dt\cld \prb - \cldh \prb + \sigma_{T}n_{e} \left(\tfrac{4}{3} \rhogam
\velb - q_{\gamma} \right) = 0.
\label{gen_eq_6}
\end{equation}

For the neutrinos, we have the moment equations
\begin{equation}
u\dt\cld q_{\nu} + \tfrac{4}{3} \theta q_{\nu} + \pi_{\nu}(\cldb)
 - \tfrac{1}{3} \cldh \rhonu = 0,
\label{gen_eq_7}
\end{equation}
\veqspc
\begin{multline}
a_{1}\dt(u\dt\dot{\cld}\dot{\pi}_{\nu}(a_{2})) +
\tfrac{4}{3} \theta a_{1}\dt\pi_{\nu}(a_{2}) + \jnul{3}(\cldb,a_{1},a_{2})
- \tfrac{1}{5} \bigl( a_{1}\dt(a_{2}\dt\cldh q_{\nu}) + a_{2}\dt(a_{1}\dt\cldh
q_{\nu}) \\
- \tfrac{2}{3} H(a_{1})\dt H(a_{2}) \cldh\dt q_{\nu} \bigr)
-\tfrac{8}{15} a_{1}\dt\sigma(a_{2}) \rhonu = 0,
\label{gen_eq_8}
\end{multline}
and for $l\geq 3$
\begin{multline}
u\dt\dot{\cld} \jnuldt{l}(a_{1},\dots,a_{l}) + \tfrac{4}{3}\theta
\jnul{l}(a_{1},\dots,a_{l}) + \jnul{l+1}(\cldb,a_{1},\dots,a_{l})
\\
- \tfrac{l}{(2l+1)l!} \bigl( a_{1}\dt\cldb \jnul{l-1}(a_{2},\dots,a_{l}) -
\tfrac{(l-1)}{(2l-1)} \jnul{l-1}(\cldb,a_{1},\dots,a_{l-2}) H(a_{l-1})
\dt H(a_{l}) + \mbox{perms} \bigr)
\\
= 0.
\label{gen_eq_9}
\end{multline}

For the photons, we have
\begin{equation}
u\dt\cld q_{\gamma} + \tfrac{4}{3} \theta q_{\gamma} + \pi_{\gamma}(\cldb)
- \tfrac{1}{3} \cldh \rho_{\gamma} =
\sigma_{T} n_{e} \left(\tfrac{4}{3} \rho_{\gamma} \velb - q_{\gamma} \right),
\label{gen_eq_10}
\end{equation}
\veqspc
\begin{multline}
a_{1} \dt (u\dt\dot{\cld}\dot{\pi}_{\gamma}(a_{2}) )
+ \tfrac{4}{3} \theta a_{1}\dt \pi_{\gamma}(a_{2}) +
\jgaml{3}(\cldb,a_{1},a_{2}) -
\tfrac{1}{5} \bigl(
a_{1}\dt(a_{2}\dt\cldh q_{\gamma}) + a_{2}\dt(a_{1}\dt\cldh q_{\gamma})
\\
- \tfrac{2}{3} H(a_{1})\dt H(a_{2}) \cldh\dt q_{\gamma} \bigr)
- \tfrac{8}{15} a_{1}\dt \sigma(a_{2}) \rho_{\gamma} =
-\tfrac{9}{10} \sigma_{T} n_{e} a_{1}\dt \pi_{\gamma}(a_{2}),
\label{gen_eq_11}
\end{multline}
and for $l\geq 3$
\begin{multline}
u\dt\dot{\cld} \jgamldt{l}(a_{1},\dots,a_{l}) + \tfrac{4}{3}\theta
\jgaml{l}(a_{1},\dots,a_{l}) + \jgaml{l+1}(\cldb,a_{1},\dots,a_{l})
\\
- \tfrac{l}{(2l+1)l!} \bigl( a_{1}\dt\cldb \jgaml{l-1}(a_{2},\dots,a_{l}) -
\tfrac{(l-1)}{(2l-1)} \jgaml{l-1}(\cldb,a_{1},\dots,a_{l-2}) H(a_{l-1})
\dt H(a_{l}) + \mbox{perms} \bigr)
\\
=  -\sigma_{T}n_{e} \jgaml{l}(a_{1},\dots,
a_{l}).
\label{gen_eq_12}
\end{multline}

The remaining equations are the propagation and constraint equations for
the kinematic variables and the electric and magnetic parts of the Weyl tensor.
The constraint equations are
\begin{equation}
\clb(a) - \qrt[(ia\wdg u\wdg\cldh)\dt\varpi + iu\wdg \cldb\wdg (a\dt\varpi)]
+ \half [iu\wdg \cldb \wdg \sigma(a) + \sigma(iu\wdg a\wdg\cldb) ] = 0
\label{gen_eq_13}
\end{equation}
\veqspc
\begin{equation}
\clb(\cldb) = -\half \kappa [(\rho + p) iu\wdg\varpi + iu\wdg\cldh\wdg q]
\label{gen_eq_14}
\end{equation}
\veqspc
\begin{equation}
\cle(\cldb) = {\textstyle \frac{1}{6}} \kappa [2\cldh\rho + 2\theta q
+ 3 \pi(\cldb)] \label{gen_eq_15} 
\end{equation}
\veqspc
\begin{equation}
\half \cldh\dt\varpi - \sigma(\cldb) +
{\textstyle \frac{2}{3}} \cldh \theta = -\kappa q \label{gen_eq_16} 
\end{equation}
\veqspc
\begin{equation}
\cldh\dt (iu\wdg \varpi) = 0, \label{gen_eq_17}
\end{equation}
where $\rho$ and $p$ are the total energy density and pressure measured by
an observer moving with velocity $u$:
\begin{align}
\rho & \equiv \rhonu + \rhogam + \rhob + \rhocdm \label{gen_eq_18}\\
p &\equiv \third \rhonu + \third \rhogam + \prb \label{gen_eq_19},
\end{align}
and $q$ and $\pi(a)$ are the total heat flux and anisotropic stress
in the frame defined by $u$:
\begin{align}
q &\equiv q_{\nu} + q_{\gamma} + (\rhob + \prb)\velb \label{gen_eq_20}\\
\pi(a) &\equiv \pi_{\nu}(a) + \pi_{\gamma}(a). \label{gen_eq_21}
\end{align}
Note that $\rho$ and $p$ are independent of the choice of $u$ to first-order,
and that there is no CDM contribution to $q$ since we have chosen $u$ equal
to the CDM velocity.

The propagation equations are
\begin{multline}
-u\dt\dot{\cld}\dot{\cle}(a) - \theta \cle(a) - \cli_{\clb}(a)
= {\textstyle \frac{1}{12}} \kappa [6(\rho + p)\sigma(a) + 3(a\dt\cldh q +
\cldb(a\dt q)) \\
- 6u\dt\dot{\cld}\dot{\pi}(a) - 2\theta \pi(a) - 2 H(a)\cldh\dt q ]
\label{gen_eq_22}   
\end{multline}
\veqspc
\begin{equation}
-u\dt\dot{\cld}\dot{\clb}(a) - \theta \clb(a) + \cli_{\cle}(a) =
- \qrt \kappa [iu\wdg\cldb\wdg\pi(a) + \pi(i u\wdg a\wdg\cldb)]
\label{gen_eq_23}
\end{equation}
\veqspc
\begin{equation}
u\dt\dot{\cld}\dot{\sigma}(a) +{\textstyle \frac{2}{3}} \theta \sigma(a) 
+ \cle(a) + \half \kappa \pi(a) = 0
\label{gen_eq_24}
\end{equation}
\veqspc
\begin{equation}
u\dt\cld \varpi +
{\textstyle \frac{2}{3}} \theta \varpi= 0, \label{gen_eq_25}
\end{equation}
where $\cli_{\clb}(a)$ is a trace-free, symmetric, spatial tensor defined by
\begin{equation}
\cli_{\clb}(a) \equiv \half [ i u\wdg \cldb \wdg \clb(a) + \clb(i u\wdg a\wdg
\cldb)],
\label{gen_eq_26}
\end{equation}
where
\begin{equation}
\clb(i u\wdg a\wdg\cldb) \equiv H[H(b)\dt\dot{\cld}
\dot{\clb}(iu\wdg a\wdg \db)],
\end{equation}
and $\cli_{\cle}(a)$ is defined similarly. There is some redundancy in
these equations. For example, equations~\eqref{gen_eq_13}, \eqref{gen_eq_16}
and the integrability condition
\begin{equation}
\cldh \wdg \cldh \theta = - \varpi u\dt\cld \theta, \label{gen_eq_27}
\end{equation}
imply the constraint~\eqref{gen_eq_14}. Similarly, equation~\eqref{gen_eq_13}
along with the propagation equations~\eqref{gen_eq_24} and~\eqref{gen_eq_25}
imply the propagation equation~\eqref{gen_eq_23}. It follows that $\clb(a)$
may be eliminated from the full set of equations by making use of the
constraint~\eqref{gen_eq_13}. This turns out to be a necessary step when the
covariant first-order quantities are harmonically expanded for scalar, vector
or tensor perturbations.

Some comments are in order regarding the equations presented in this section.
The equations are both covariant and gauge-invariant (in the sense of
being independent of any map between the lumpy universe and the FRW model).
Gauge-invariance ensures that the gauge-problems that have plagued other
approaches are not present, while covariance ensures that we are working
with quantities
which are straightforward to interpret physically. The equations describe
scalar, vector and tensor perturbations in a unified manner, and are
independent of any harmonic analysis into spatial modes. Furthermore,
we have not had to specify the background FRW model yet (there is
an implicit assumption that the universe is approximately FRW when
the first-order, covariant variables are constructed).

\subsection{Scalar Perturbations in a {$K=0$} Universe}

In this section we reduce the general equations of the previous section
to a set of equations for scalar-valued, first-order gauge-invariant
variables describing the evolution of the
density inhomogeneities and the CMB anisotropy for scalar perturbations
in a universe which is approximately a $K=0$ FRW model. These equations
thus describe the standard CDM model in a covariant and gauge-invariant manner.
The equations split into a set of algebraic
relations (from the constraint equations) and propagation equations for
scalar variables. The moment equations for the photon and neutrino
distributions (for $l\geq 3$) are equivalent to those given
elsewhere~\cite{ma95}, where a Fourier expansion of the spatial
variation, and a Legendre expansion of the angular dependence of the
CMB anisotropy are made.

Scalar perturbations may be characterised in a covariant manner by demanding
that the magnetic part of the Weyl tensor and the vorticity both vanish
identically. The vanishing of the vorticity ensures that $u$ is a hypersurface
orthogonal vector field.
The perturbation equations place strong restrictions on the remaining
non-zero variables, which may be satisfied by constructing the vector and
tensor variables from spatial derivatives of the scalar (harmonic)
eigenfunctions $\qk$ of the generalised Helmholtz equation:
\begin{equation}
\cldhsq \qk = \tfrac{k^{2}}{S^{2}} \qk,
\label{scal_eq_1}
\end{equation}
which are constructed to satisfy
\begin{equation}
u\dt\cld \qk = \ord(1).
\label{scal_eq_2}
\end{equation}
From the $\qk$ we form a scalar valued tensor
$\qkl{1}(a)$:
\begin{equation}
\qkl{1}(a) \equiv \tfrac{S}{k} a\dt\cldh \qk,
\label{scal_eq_3}
\end{equation}
which has the properties
\begin{equation}
\qkl{1}(u) = 0, \qquad u\dt\dot{\cld}\qkldt{1}(a) = \ord(1).
\label{scal_eq_4}
\end{equation}
We then define scalar-valued tensors $\qkl{l}(a_{1},\dots,a_{l})$ by
the recursion formula (for $l>1$)
\begin{multline}
\qkl{l}(a_{1},\dots,a_{l}) = \tfrac{1}{l!}\tfrac{S}{k} \bigl(
a_{1}\dt\cldb \qkl{l-1}(a_{2},\dots,a_{l}) - \tfrac{l-1}{2l-1}
H(a_{1})\dt H(a_{2}) \qkl{l-1}(\cldb,a_{3},\dots,a_{l}) \\
+ \mbox{perms} \bigr).
\label{scal_eq_5}
\end{multline}
These tensors satisfy the properties
\begin{align}
\qkl{l}(a_{1},\dots,a_{j},\dots,a_{k},\dots,a_{l}) & =
\qkl{l}(a_{1},\dots,a_{k},\dots,a_{j},\dots,a_{l}) \notag \\
\qkl{l}(u,a_{2},\dots,a_{l}) & = 0  \notag \\
\qkl{l}(\db,b,a_{3},\dots,a_{l}) & = 0 \notag \\
u\dt\dot{\cld}\qkldt{l}(a_{1},\dots,a_{l}) & = \ord(1),
\end{align}
which are readily proved by induction.
We shall also make use of the vector $\qkv \equiv \da\qkl{1}(a)$ and the
vector-valued tensor $\qk(a) \equiv \db \qkl{2}(a,b)$.

The definitions above are independent of the spatial curvature $K$ of the
background model. However, the differential properties of the $\qkl{l}(a_{1},
\dots,a_{l})$ are dependent on the value of $K$. Some of these properties are
listed in the appendix to Bruni~\etal~\cite{bruni92}\ The
only new result we require is
\begin{equation}
\qkl{l}(\cldb,a_{2},\dots,a_{l}) = \tfrac{l}{2l-1} \tfrac{k}{S}
\qkl{l-1}(a_{2},\dots,a_{l}),
\label{scal_eq_10}
\end{equation}
which is valid for $K=0$ only.

We separate out the spatial and temporal dependence of the covariant,
first-order variables by expanding in the harmonics derived from the
$\qk$ in the following manner:
\begin{alignat}{2}
\clx_{i} & = \sum_{k} k \clx_{i k} \qkv, & \qquad
\clz & = \sum_{k} \tfrac{k^{2}}{S} \clz_{k} \qkv \label{scal_eq_11} \\
\cle(a) & = \sum_{k} \left(\tfrac{k}{S}\right)^{2} \Phi_{k} \qk(a), & \qquad
\sigma(a) & = \sum_{k} \left( \tfrac{k}{S} \right) \sigma_{k} \qk(a)
\label{scal_eq_12} \\
q_{\nu} & = \rhonu \sum_{k} \qnuk \qkv, & \qquad
q_{\gamma} & = \rhogam \sum_{k} \qgamk \qkv \label{scal_eq_13} \\
\pi_{\nu}(a) & = \rhonu \sum_{k} \pinuk \qk(a), & \qquad 
\pi_{\gamma}(a) & = \rhogam \sum_{k} \pigamk \qk(a) \label{scal_eq_14}.
\end{alignat}
We also expand $\velb$ as
\begin{equation}
\velb = \sum_{k} \vk \qkv.
\label{scal_eq_15}
\end{equation}
Finally, we assume that the higher-order moments of the (energy-integrated)
neutrino and photon distribution functions may also be expanded in harmonics.
By considering the zero-order form of the scalar harmonics $\qk$, it is
straightforward to show that this assumption is equivalent to that which is
usually made~\cite{ma95} (the
Fourier components of the distribution function are axisymmetric about the
wavevector $\bk$). The procedure we have adopted here, of expanding the
angular dependence of the distribution function in symmetric,
traceless, covariant spatial tensors, which are then expanded in the
appropriate harmonic tensors, derived from the scalar harmonics, should be
compared to the usual approach of Fourier expanding (for $K=0$) all
variables, and then performing a Legendre expansion of the angular
dependence of the distribution function modes about the wavevector $\bk$.
Although ultimately equivalent, the approach taken here has the advantage
that it goes over unchanged to the case of tensor perturbations (although
it is now $\qk^{(2)}(a_{1},a_{2})$ which satisfies the generalised Helmholtz
equation).
For later convenience, we expand the angular moments of the
distribution function in the form
\begin{align}
\jnul{l}(a_{1},\dots,a_{l}) & = \rhonu \sum_{k} \jnulk{l} \qkl{l}(a_{1},\dots,
a_{l}) \label{scal_eq_16} \\
\jgaml{l}(a_{1},\dots,a_{l}) & = \rhogam \sum_{k} \jgamlk{l} \qkl{l}(a_{1},
\dots,a_{l}) \label{scal_eq_17}.
\end{align}

It is now a simple matter to substitute the harmonic expansions of the
covariant variables into the
general equations of Section~\ref{sec_gen} (with $\clb(a)$ and $\varpi$ set
to zero), to obtain equations for the scalar expansion coefficients.
We assume that the variations in baryon pressure $\prb$ due to entropy
variations are insignificant compared to those due to variations in
$\rhob$. Accordingly, we write
\begin{equation}
\cld \prb = \csound^{2} \cld \rhob,
\label{scal_eq_18}
\end{equation}
where $\csound$ is the adiabatic sound speed in the baryon/electron fluid (this
is different from the sound speed in the coupled baryon/photon plasma). 

With this assumption we obtain the following equations for scalar perturbations
in a $K=0$ universe: the spatial gradients of the densities give
\begin{align}
\clxnudtk & = -\tfrac{4}{3} \tfrac{k}{S} \clz_{k} - \tfrac{k}{S} \qnuk
\label{scal_eq_19} \\
\clxgamdtk & = -\tfrac{4}{3} \tfrac{k}{S} \clz_{k} - \tfrac{k}{S} \qgamk
\label{scal_eq_20} \\
\clxbdtk & = - \left(1 + \tfrac{\prb}{\rhob}\right)\tfrac{k}{S}(\clz_{k} + \vk)
+ \left(\tfrac{\prb}{\rhob} - \csound^{2}\right) \theta \clxbk
\label{scal_eq_21}\\
\clxcdmdtk & = - \tfrac{k}{S} \clz_{k}, \label{scal_eq_22}
\end{align}
(an overdot on a scalar denotes $u\dt\cld$), the spatial gradient of $\theta$
gives
\begin{equation}
\tfrac{k}{S} \dot{\clz}_{k} + \third \tfrac{k}{S} \theta \clz_{k} +
\half \kappa \bigl( 2\rhogam\clxgamk + 2\rhonu\clxnuk + (1+3\csound^{2})
\rhob\clxbk + \rhocdm\clxcdmk \bigr) = 0, \label{scal_eq_23}
\end{equation}
the heat fluxes give
\begin{align}
\qnudtk + \tfrac{2}{3}\tfrac{k}{S} \pinuk - \tfrac{1}{3}\tfrac{k}{S} \clxnuk
& = 0 \label{scal_eq_24} \\
\qgamdtk + \tfrac{2}{3}\tfrac{k}{S}\pigamk - \tfrac{1}{3}\tfrac{k}{S} \clxgamk & =
\sigma_{T}n_{e}\left(\tfrac{4}{3} \vk - \qgamk \right) \label{scal_eq_25} \\
\left(1+\tfrac{\prb}{\rhob}\right)\bigl(\dot{v}_{k} + \tfrac{1}{3}(1-3
\csound^{2}) \theta \vk \bigr) - \csound^{2}\tfrac{k}{S} \clxbk & =
- \tfrac{\rhogam}{\rhob} \sigma_{T} n_{e} \left(\tfrac{4}{3} \vk -
\qgamk\right),
\label{scal_eq_26}
\end{align}
the remaining moment equations are, for $l\geq 3$
\begin{align}
\pinudtk + \tfrac{3}{5}\tfrac{k}{S} \jnulk{3} - \tfrac{2}{5} \tfrac{k}{S}
\qnuk - \tfrac{8}{15} \tfrac{k}{S} \sigma_{k} & = 0 \label{scal_eq_27} \\
\pigamdtk + \tfrac{3}{5}\tfrac{k}{S} \jgamlk{3} - \tfrac{2}{5} \tfrac{k}{S}
\qgamk - \tfrac{8}{15} \tfrac{k}{S} \sigma_{k} &= -\tfrac{9}{10} \sigma_{T}
n_{e} \pigamk \label{scal_eq_28} \\
\jnuldtk{l} + \tfrac{k}{S} \left( \tfrac{l+1}{2l+1} \jnulk{l+1}
- \tfrac{l}{2l+1} \jnulk{l-1} \right) &= 0
\label{scal_eq_29} \\
\jgamldtk{l} + \tfrac{k}{S} \left( \tfrac{l+1}{2l+1} \jgamlk{l+1}
- \tfrac{l}{2l+1} \jgamlk{l-1} \right) &= - \sigma_{T} n_{e} \jgamlk{l}
\label{scal_eq_30},
\end{align}
the electric part of the Weyl tensor gives
\begin{multline}
-\left(\tfrac{k}{S}\right)^{2} \left(\dot{\Phi}_{k} + \tfrac{1}{3} \theta
\Phi_{k} \right) = \half \kappa \tfrac{k}{S} \bigl(
(\rho+p)\sigma_{k} + \rhonu\qnuk + \rhogam\qgamk + (\rhob + \prb)\vk \bigr) \\
+ \half \kappa \theta (\rhonu\pinuk + \rhogam\pigamk) - \half\kappa
(\rhonu \pinudtk + \rhogam \pigamdtk ),
\label{scal_eq_31}
\end{multline}
and the shear tensor gives
\begin{equation}
\tfrac{k}{S} \left( \dot{\sigma}_{k} + \third \theta \sigma_{k} \right)
+ \left(\tfrac{k}{S}\right)^{2} \Phi_{k} + \half\kappa (\rhonu \pinuk
+ \rhogam \pigamk) = 0.
\label{scal_eq_32}
\end{equation}
Finally, the harmonic expansions of the constraint equations give
\begin{multline}
2\left(\tfrac{k}{S}\right)^{3} \Phi_{k} - \kappa \tfrac{k}{S} \bigl(
\rhonu(\clxnuk + \pinuk) + \rhogam(\clxgamk + \pigamk) + \rhocdm\clxcdmk
+ \rhob \clxbk \bigr)
\\
- \kappa \theta \bigl( \rhonu \qnuk + \rhogam \qgamk + (\rhob+\prb)\vk \bigr)
= 0, \label{scal_eq_33}
\end{multline}
and
\begin{equation}
\tfrac{2}{3} \left(\tfrac{k}{S}\right)^{2} (\clz_{k} - \sigma_{k}) +
\kappa \bigl( \rhonu\qnuk + \rhogam\qgamk + (\rhob + \prb)\vk \bigr) = 0.
\label{scal_eq_34}
\end{equation}
It is straightforward to show that these constraint equations are preserved by
the propagation equations.

These gauge-invariant equations should be compared to the (gauge-dependent)
equations used in most calculations of the CMB anisotropy (see, for
example Ma and Bertshinger~\cite{ma95}).
Note that the equations for $\jgamlk{l}$ and
$\jnulk{k}$ for $l\geq 3$ are equivalent to those usually found in the
literature since the moments of the distribution function in such
gauge-dependent calculations are actually gauge-invariant for $l\geq 3$.

\subsection{The CMB temperature anisotropy}

The equations of the previous section may be solved numerically for appropriate
initial conditions~\cite{chall97-cmb} to give
the $\jgamlk{l}$ at the present epoch. These moments fully describe the
CMB temperature anisotropy. Denote the full sky average of the CMB temperature
by $T_{0}$, and the (gauge-invariant) temperature difference from the mean
along a spatial direction $e$ by $\delta T(e)$. Then we have~\cite{ellis94er}
\begin{equation}
\frac{(T_{0} + \delta T(e))^{4}}{T_{0}^{4}} =
\frac{\sum_{l=0}^{\infty} \int dE\, E^{3} \fgaml{l}(e,\dots,e)}
{\int dE\, E^{3} \fgaml{0}},
\label{scal_eq_35}
\end{equation}
so that, to first-order,
\begin{equation}
4\frac{\delta T(e)}{T_{0}} = \frac{4\pi}{\rhogam} 
\sum_{l=1}^{\infty} \int dE\, E^{3} \fgaml{l}(e,\dots,e).
\label{scal_eq_36}
\end{equation}
Recalling the definition~\eqref{be_eq_13}, we find
\begin{equation}
\frac{\delta T(e)}{T_{0}} = -\frac{3}{4}
\frac{q_{\gamma}\dt e}
{\rhogam} + \frac{15}{8} \frac{\pi_{\gamma}(e)\dt e}{\rhogam}
+ \frac{1}{4\rhogam}\sum_{l=3}^{\infty} \frac{(2l+1)(2l)!}{(-2)^{l} (l!)^{2}}
\jgaml{l}(e,\dots,e).
\label{scal_eq_37}
\end{equation}
The right-hand side of~\eqref{scal_eq_37} is the covariant harmonic expansion
of the temperature anisotropy. The temperature anisotropy may be expanded
in the more familiar spherical harmonics by introducing a covariant triad,
orthogonal to $u$, at the observation point. We now have all the ingredients
needed to calculate the CMB anisotropy for scalar perturbations in a flat
CDM model, for given initial conditions. The numerical calculation of
the anisotropy for such a model are underway, and the results will be reported
elsewhere~\cite{chall97-cmb}. Our results should confirm the
calculations of other groups, once the gauge-invariant information is
extracted from their results (the CMB power spectrum, $C_{l}$,
for $l>0$).

\subsection{Large Scale Anisotropies}

On large angular scales, the dominant contributions to the anisotropy may be
extracted analytically without recourse to a full numerical treatment.
In this section, we show how this can be done within the
framework developed above. Our gauge-invariant treatment of large-scale
anisotropies is complementary to the discussion by
Ellis and Dunsby~\cite{ellis-er97} elsewhere in this volume, who use
a two-fluid model for the matter and radiation after recombination, rather
than the kinetic theory approach adopted here.

For semi-analytic work, it is convenient to use the Boltzmann equation
in the form of equation~\eqref{be_eq_26}. Defining
\begin{equation}
\delta_{T}(e) \equiv \frac{\delta T(e)}{T_{0}},
\label{ls_eq_1}
\end{equation}
where $\delta T(e)$ is the gauge-invariant temperature fluctuation from the
mean (see equation~\eqref{scal_eq_35}), we can write~\eqref{be_eq_26} in the
form
\begin{multline}
\delta_{T}(e)'+ \sigma_{T} n_{e} \delta_{T}(e) =
e\dt\sigma(e) + e\dt w - \frac{1}{3}\theta(1+ 4 \delta_{T}(e))
 - \frac{\rho_{\gamma}'}{4\rho_{\gamma}} (1 + 4\delta_{T}(e))
\\
-\sigma_{T} n_{e} \left(e\dt v_{b} - \frac{3}{16} \frac{e\dt\pi_{\gamma}(e)}
{\rho_{\gamma}}\right),
\label{ls_eq_2}
\end{multline}
where a prime denotes differentiation with respect to the parameter $\lambda$
along the geodesic, where, for this section, we take $(u+e)\dt \cld \lambda=1$.
Equation~\eqref{ls_eq_2} is correct to first-order.
Note that in writing~\eqref{ls_eq_2} we have not made a physical choice for
the velocity $u$ yet, so we allow the possibility that $w$ does not vanish.
(In the previous section we took $u$ to coincide with the velocity of the CDM,
but we relax that restriction here.)
Writing the Boltzmann equation in the form~\eqref{ls_eq_2}
is useful since the equation may
be integrated along the geodesic to determine the temperature anisotropy along
the given direction on the sky. Before doing this, it is convenient to
eliminate $\theta$ from~\eqref{ls_eq_2}. In this section we shall
consider a model with interacting baryons and radiation only, so that we
describe the same situation as Ellis and Dunsby elsewhere in this
volume~\cite{ellis-er97} (see also Dunsby~\cite{dunsby96b}).
In this case, it is
convenient to define $u$ to be equal to the velocity of the baryons, so
that $v_{b}=0$. Using~\eqref{bar_eq_3}, and neglecting baryon pressure, we find
\begin{equation}
\theta = - \frac{\rho_{b}'}{\rho_{b}} + \frac{e\dt\cldh \rho_{b}}{\rho_{b}},
\label{ls_eq_3}
\end{equation}
and using~\eqref{bar_eq_4}, we may write the acceleration as
\begin{equation}
w = \sigma_{T} n_{e} \frac{q_{\gamma}}{\rho_{b}},
\label{ls_eq_4}
\end{equation}
which shows that after recombination, the acceleration is negligible,
particularly in a universe which is matter dominated at recombination.
Substituting into~\eqref{ls_eq_2}, we find
\begin{equation}
\delta_{T}(e)'+ \sigma_{T} n_{e} \delta_{T}(e) =
e\dt\sigma(e) - \frac{1}{3} \frac{e\dt\cldh \rho_{b}}{\rho_{b}} +
\frac{\rho_{b}'}{3\rho_{b}} - \frac{\rho_{\gamma}'}{4\rho_{\gamma}}
+ \sigma_{T} n_{e} \left( \frac{e\dt q_{\gamma}}{\rho_{b}} + \frac{3}{16}
\frac{e\dt\pi_{\gamma}(e)}{\rho_{\gamma}} \right).
\label{ls_eq_5}
\adjusttag{+8pt}
\end{equation}
This equation may be integrated up the null geodesic from some point in the
distant past (where $\lambda=\lambda_{i}$) to the reception point $R$ (where
$\lambda=\lambda_{R}$). Introducing the optical depth $\kappa(\lambda)$ along
the line of sight, defined by
\begin{equation}
\kappa(\lambda) = \int_{\lambda}^{\lambda_{R}} n_{e} \sigma_{T} \, d\lambda,
\label{ls_eq_6}
\end{equation}
we find that~\eqref{ls_eq_5} integrates to give
\begin{multline}
\bigl(\delta_{T}(e)\bigr)_{R} = \int_{\lambda_{i}}^{\lambda_{R}} -\kappa'\et{-\kappa}
\Bigl[\frac{e\dt q_{\gamma}}{\rho_{b}} + \frac{3}{16}
\frac{e\dt\pi_{\gamma}(e)}{\rho_{\gamma}} - \frac{1}{3} \ln \rho_{b}
+ \frac{1}{4} \ln \rho_{\gamma}  \Bigr]
\\
+ \et{-\kappa} \Bigl[ e\dt\sigma(e) - \frac{1}{3}
\frac{e\dt\cldh \rho_{b}}{\rho_{b}} \Bigr] \, d\lambda,
\label{ls_eq_7}
\end{multline}
where we have integrated by parts, we have assumed that terms evaluated at
$\lambda_{i}$ are negligible due to the large optical depth, and we have
neglected a direction-independent (monopole) term which must be
cancelled by other terms in the integral.
The notation $(A)_{R}$ denotes the quantity $A$ evaluated at the
point $R$.
An expression similar
to~\eqref{ls_eq_7} forms the starting point of the `line of sight integration
approach' to calculating CMB anisotropies~\cite{seljak96}, but note
that~\eqref{ls_eq_7} is true for scalar, vector and tensor modes in
almost FRW universes with any value of $K$, and has not assumed any Fourier
decomposition. Furthermore, the expression only involves physically-defined
gauge-invariant variables, such as the fractional temperature fluctuation from
the mean $\delta_{T}(e)$.

The quantity $-\kappa' \et{-\kappa}$ defines the visibility function, which is
the probability density that a photon was last scattered at $\lambda$.
The visibility function peaks at a redshift $z\simeq 1100$ and has a dispersion
$\simeq 70$ in redshift~\cite{jones85}. It follows that on angular scales
larger than $8'\Omega^{-1/2}$, the visibility function may be approximated
by a $\delta$-function, whose support defines the last scattering surface.
With this approximation, $\et{-\kappa}=0$ before last scattering
(tight-coupling), and $\et{-\kappa}=1$ after last scattering (free-streaming),
so that equation~\eqref{ls_eq_7} reduces to
\begin{equation}
\bigl(\delta_{T}(e)\bigr)_{R} =
\Bigl(\frac{e\dt q_{\gamma}}{\rho_{b}} + \frac{3}{16}
\frac{e\dt\pi_{\gamma}(e)}{\rho_{\gamma}} - \frac{1}{3} \ln \rho_{b}
+ \frac{1}{4} \ln \rho_{\gamma}  \Bigr)_{A}
+ \int_{\lambda_{A}}^{\lambda_{R}}\Bigl[ e\dt\sigma(e) - \frac{1}{3}
\frac{e\dt\cldh \rho_{b}}{\rho_{b}} \Bigr] \, d\lambda,
\label{ls_eq_8}
\adjusttag{+5pt}
\end{equation}
where $A$ is the point where the geodesic intersects the last
scattering surface. Taking the difference between two directions on the sky, we
find the following expression for the gauge-invariant
temperature difference $\Delta T$
\begin{multline}
\left( \frac{\Delta T}{T_{0}}\right)_{R} = \frac{1}{4} \Delta
(\ln \rho_{\gamma})_{E} - \frac{1}{3} \Delta(\ln \rho_{b})_{E} +
\frac{\Delta(e\dt q_{\gamma})_{E}}{\rho_{b}} + \frac{3}{16}
\frac{\Delta(e\dt\pi_{\gamma}(e))_{E}}{\rho_{\gamma}}
\\
+
\Delta  \int\Bigl[ e\dt\sigma(e) - \frac{1}{3}
\frac{e\dt\cldh \rho_{b}}{\rho_{b}} \Bigr] \, d\lambda,
\label{ls_eq_9}
\end{multline}
where, for example,
\begin{equation}
\Delta(\ln\rho_{\gamma})_{E} \equiv (\ln\rho_{\gamma})_{A} -
(\ln\rho_{\gamma})_{B},
\label{ls_eq_10}
\end{equation}
with $B$ the point of intersection of the other null geodesic with the
last scattering surface, and
\begin{equation}
\Delta \int (\ ) \, d\lambda \equiv \int_{\lambda_{A}}^{\lambda_{R}} (\ )
\, d\lambda - \int_{\lambda_{B}}^{\lambda_{R}} (\ ) \, d\lambda.
\label{ls_eq_11}
\end{equation}
The first term on the right-hand side of~\eqref{ls_eq_9} is just the
fractional difference in radiation temperature between the points of
emission $A$ and
$B$ on the last scattering surface, while the second term describes the
effects of inhomogeneity in the baryon density. It is convenient to
write the sum of these two terms as a line integral in the last scattering
surface. Denoting by $\dxcov$ the covariant element of length for an
arbitrary path connecting $A$ and $B$, we have
\begin{align}
\frac{1}{4}\Delta(\ln \rho_{\gamma})_{E} -
\frac{1}{3}\Delta(\ln \rho_{b})_{E} & =
\int_{B}^{A} \left[ \frac{\cld \rho_{\gamma}}{4\rho_{\gamma}} -
\frac{\cld \rho_{b}}{3\rho_{b}} \right] \dt \dxcov \notag \\
& =\int_{B}^{A} \left[ \left( \frac{\cldh \rho_{\gamma}}{4\rho_{\gamma}}
- \frac{\cldh \rho_{b}}{3\rho_{b}} \right) + u
\left(\frac{u\dt\cld \rho_{\gamma}}{4\rho_{\gamma}} -
\frac{u\dt\cld \rho_{b}}{3\rho_{b}} \right) \right]\dt \dxcov \notag \\
& = \int_{B}^{A} [ \cls_{\gamma b} - \tfrac{1}{3} u \cldh\dt v_{\gamma}]
\dt \dxcov,
\label{ls_eq_12}
\end{align}
where
\begin{equation}
\cls_{\gamma b} \equiv \frac{\cldh \rho_{\gamma}}{4\rho_{\gamma}}
- \frac{\cldh \rho_{b}}{3\rho_{b}}, \qquad
v_{\gamma} \equiv \frac{3q_{\gamma}}{4\rho_{\gamma}}.
\label{ls_eq_13}
\end{equation}
The covariant vector $\cls_{\gamma b}$ is a covariant measure of the
entropy perturbations (see, for example, Bruni~\etal~\cite{bruni92}), and
$v_{\gamma}$ is the effective radiation velocity (relative to the
baryon frame). If we take the integration path to lie entirely in the
last scattering surface, $u\dt \dxcov = \ord(1)$, so that~\eqref{ls_eq_12}
reduces to
\begin{equation}
\frac{1}{4}\Delta(\ln \rho_{\gamma})_{E} -
\frac{1}{3}\Delta(\ln \rho_{b})_{E} = \int_{B}^{A}
\cls_{\gamma b} \dt \dxcov,
\label{ls_eq_14}
\end{equation}
which shows that the contribution of the first two terms
in~\eqref{ls_eq_9} depends only on the entropy perturbations on the last
scattering surface. In particular, for perturbations which are exactly
adiabatic at last scattering, there is no contribution from these terms.
The third term
in~\eqref{ls_eq_9} is a Doppler
term which is negligible for a universe which is matter dominated at 
recombination, and the fourth arises from viscous effects in the photons.
Before recombination, both $q_{\gamma}$ and $\pi_{\gamma}$
are first-order in the photon mean free time $(n_{e}\sigma_{T})^{-1}$ (recall
that we have defined $u$ to coincide with the baryon velocity), so that
in the tight-coupling/instantaneous recombination approximation these terms
may be ignored. It follows that for perturbations which are exactly
adiabatic at last scattering, only the integral term on the right-hand side
of~\eqref{ls_eq_9} remains.

An expression similar to~\eqref{ls_eq_9} is derived elsewhere in this volume
by Ellis and Dunsby~\cite{ellis-er97}, although they used a two fluid approach
rather than the kinetic theory approach adopted here. To make contact with
their work, we consider a universe which is already matter dominated at
recombination. In this case, the baryon velocity coincides with the velocity
of the energy frame (the frame in which the heat flux vanishes) which was
used by Ellis and Dunsby, and we may replace $\rho_{b}$ by the total density
$\rho$ in the integral in~\eqref{ls_eq_8}. Furthermore, we ignore the
terms involving $q_{\gamma}$ and $\pi_{\gamma}$, as discussed above, to
obtain
\begin{equation}
\bigl(\delta_{T}(e)\bigr)_{R} = \frac{1}{4} (\ln \rho_{\gamma})_{A}
- \frac{1}{3} (\ln \rho_{b})_{A} + \int_{\lambda_{A}}^{\lambda_{R}}
\left[ e\dt \sigma(e) - \frac{1}{3} \frac{e\dt\cldh \rho}{\rho}\right]
\, d\lambda.
\label{ls_eq_15}
\end{equation}
For ease of comparison we also give the corresponding expression from Ellis and
Dunsby~\cite{ellis-er97} in our notation:
\begin{equation}
\bigl(\delta_{T}(e)\bigr)_{R} = \cla + \int_{\lambda_{A}}^{\lambda_{R}}
\left[ e\dt \sigma(e) - \frac{1}{3} \frac{e\dt\cldh \rho}{\rho}\right]
\, d\lambda ,
\label{ls_eq_16}
\end{equation}
where
\begin{equation}
\cla \equiv - \int_{\lambda_{A}}^{\lambda_{R}} e\dt (\cls_{\gamma b} -
\tfrac{1}{3} \cldh \dt v_{\gamma} ) \, d\lambda.
\label{ls_eq_17}
\end{equation}
By making use of the rearrangement in~\eqref{ls_eq_12}, and noting that
$\dxcov = (e+u) d\lambda$ along the geodesic, we find that
$\cla$ reduces to the first two terms on the right-hand side
of~\eqref{ls_eq_15} (after subtraction of a monopole term).
It follows that the expression for the temperature anisotropy in
Ellis and Dunsby
is in agreement with that derived here, under the assumptions outlined above.
Note that the part of $\cla$ which is significant observationally is
determined only by the entropy perturbation at last scattering, and so this
term will vanish for perturbations which are adiabatic at last scattering,
even though the perturbations do not remain adiabatic after decoupling.

\subsection{Conclusion}

The gauge-invariant calculation of CMB anisotropies may be performed in a
fully covariant manner with the methods outlined in this section. The equations
of Section~\ref{sec_gen} are a complete description of the 
evolution of perturbations
in a CDM universe. They include all types of perturbation
(scalar, vector and tensor) implicitly and are independent of the
curvature of the
background FRW model. We have demonstrated how the equation set may be reduced
to scalar equations for the case of scalar perturbations in a $K=0$ universe.
The extension to $K\neq 0$ and vector or tensor modes is straightforward.
The results of a numerical solution of the equations for scalar perturbations
will be given elsewhere~\cite{chall97-cmb}. We expect that these
calculations will confirm the result of other groups who have made their
calculations by working carefully in specific gauges.

\section{CMB Anisotropies in Global Defect Theories}
\label{sec_defs}

Our final topic is a review of the results of some recent calculations by Pen,
Seljak and Turok~\cite{pen97,seljak97b} of the power spectra in global
defect theories. Unlike the calculation of CMB anisotropies in
inflationary models, the calculation for defect theories has been plagued
by computational difficulties. These difficulties arise from the continual
generation of perturbations by the causal source (the defects). One must
solve for the non-linear evolution of the source, as well as the linearised
response of the Einstein/fluid/Boltzmann equations to the causal sources.

\begin{figure}[t!]
\begin{center}
 
 
\epsfig{figure=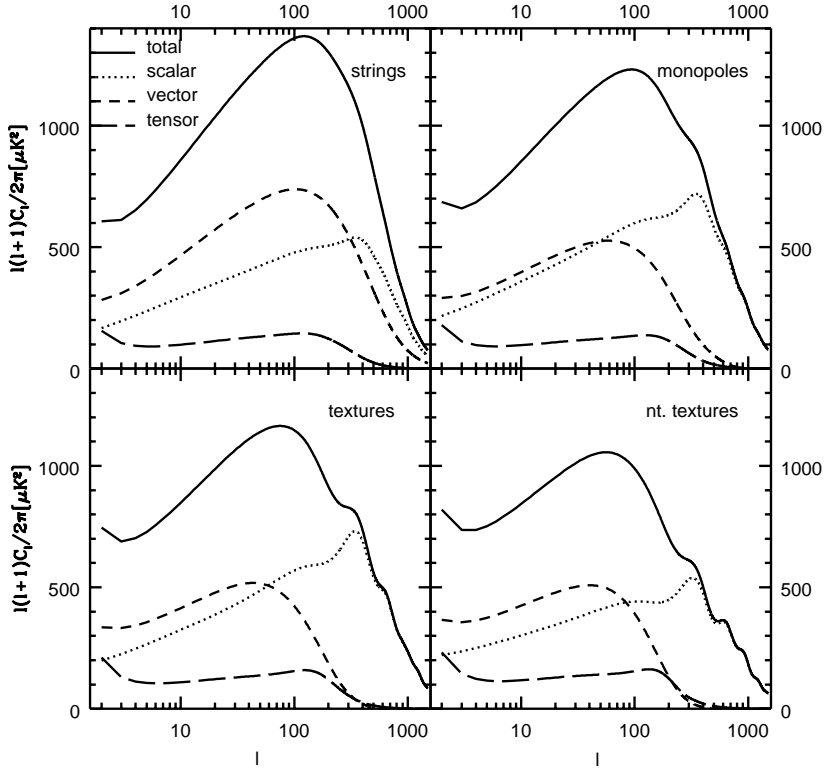,angle=0,width=11cm}

\end{center}
\caption{The contributions to the CMB power spectrum from scalar, vector and
tensor components in global defect models with strings, monopoles, textures
and non-topological textures. Reproduced with permission from Pen, Seljak and
Turok (1997). }
\label{fig_defs_1}
\end{figure}

Recently, an efficient technique has been developed for the calculation of
power spectra in defect theories~\cite{pen97}, which makes use of the
unequal time
correlator of the defect stress-energy tensor.
This technique has provided the
first accurate computations of the CMB and matter power spectra in
global defect theories. The results of the calculation for the
CMB power spectra (the $C_{l}$) are shown in Figure~\ref{fig_defs_1}, for
global defects comprised of strings, monopoles, textures and non-topological
textures. The contributions from scalar, vector and tensor modes are shown
separately, as well as their total. The most striking feature of the figure
is the large contribution from vector modes to the power on large angular
scales. The vector modes dominate the total signal for $l \simlt 100$, at
which point they are suppressed by the horizon size at recombination.
Vector modes are insignificant in inflationary models since the vorticity
decays as the universe expands to conserve angular momentum. However, in defect
theories, the defects are a continual source of vorticity giving rise to a
significant vector component to the CMB power spectrum. In addition, the
continual (causal) sourcing of the fluid perturbations leads to
decoherence, where
scalar modes of a given scale (inside the sound horizon) oscillate with
different phases in different parts of the universe, compared to the coherent
oscillations which occur in inflationary models. This decoherence manifests
itself as the suppression of the secondary Doppler peaks in
Figure~\ref{fig_defs_1}.

\begin{figure}[t]
\begin{center}
 
 
\epsfig{figure=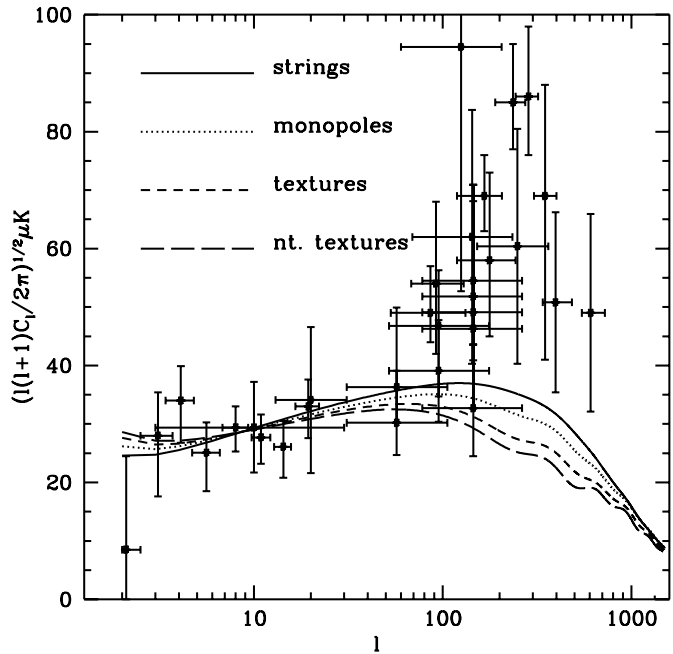,angle=0,width=9cm}

\end{center}
\caption{A comparison of the CMB power spectra calculated in defect models
with the current observations. The calculated curves are normalised to
COBE at $l=10$. Reproduced with permission from Pen, Seljak and
Turok (1997). }
\label{fig_defs_2}
\end{figure}

\begin{figure}[t]
\begin{center}
\begin{picture}(250,250)
 
 
\put(0,0){\hbox{\epsfig{figure=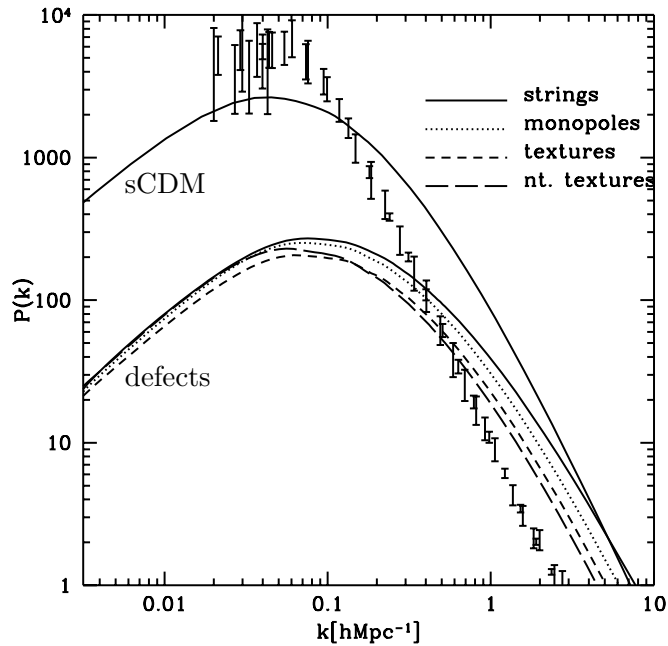,angle=0,width=9cm}}}
\put(45,180){\makebox(0,0)[l]{\small sCDM}}
\put(45,107){\makebox(0,0)[l]{\small defects}}

\end{picture}
\end{center}
\caption{The matter power spectra computed in COBE normalised defect theories
(lower curves) compared with the predictions of the standard CDM (sCDM)
model (upper curve) for the same normalisation.
The data points show the power spectrum inferred
from galaxy distributions. Reproduced with permission from Pen, Seljak and
Turok (1997). }
\label{fig_defs_3}
\end{figure}

The significant contribution of the vector modes to the large scale
power in the CMB spectrum leads to a suppression in the total signal
for $l\geq 100$ compared to inflationary models, when both are normalised at
large angular scales. In Figure~\ref{fig_defs_2} we show a comparison of the
predictions of global defect theories, with COBE normalisation at $l=10$,
to the current generation of CMB measurements. The effect of vector modes and
decoherence is to leave the predictions of the defect models systematically
below the current degree-scale data. A similar effect occurs with the matter
power spectrum, $P(k)$. The matter power spectrum calculated in
global defect theories is compared with the predictions of a standard cold
dark matter model (sCDM, with $H_{0}=50\kmsmpc$ and a scale-invariant
spectrum of initial perturbations) in Figure~\ref{fig_defs_3}.
The defect calculations and the sCDM calculations are normalised to
COBE, and are compared with the matter power spectrum inferred from the
galaxy distribution. The large contribution of vector modes to the CMB power
spectrum in the range of the COBE normalisation, suppresses the scalar
component compared to sCDM. The COBE normalised defect models predict too
little power in the matter spectrum on larger scales, and too much power
on smaller scales. As is well known, the sCDM model fares much better on
large scales, but also predicts too much small scale power.

To conclude, current degree-scale observations of the CMB power spectrum, and
the matter power spectrum inferred from galaxy clustering, do not
favour global defect theories of structure formation, on the basis of the
first accurate computations of the spectra in these theories.
The significant contribution to large scale power
from vector modes, which occurs in all defect models, leads to systematically
low (COBE normalised) predictions for degree-scale CMB observations, and
for the matter power spectrum.

\section*{Acknowledgements}

We thank U. Pen, U. Seljak and N. Turok for permission to reproduce
Figures~\ref{fig_defs_1}--\ref{fig_defs_3}, and G. Ellis and P. Dunsby for
allowing us access to the preprint of their contribution to this volume.

\end{document}